\documentclass[preprintnumbers,prd,showpacs,floatfix,preprintnumbers,amsmath,amssymb,nofootinbib,superscriptaddress]{revtex4}
\usepackage{xcolor}
\usepackage{graphicx,color,bm}

\usepackage{amsmath,amssymb}
\usepackage{epstopdf}
\usepackage{ulem}
\usepackage{array}
\usepackage{verbatim}
\usepackage[utf8]{inputenc}
\usepackage{rotating}
\newcolumntype{K}[1]{>{\centering\arraybackslash}m{#1}}

\newcommand{\Mpl}{M_{\textrm{Pl}}}

\def\al{\alpha}

\def\S{\mathcal{S}}

\def\m{\mathrm{m}}

\usepackage[colorlinks,linkcolor=blue,citecolor=blue]{hyperref}

\usepackage{soul}

\usepackage{booktabs,multirow,array}

\newcommand{\orcid}[1]{\href{https://orcid.org/#1}{\,\includegraphics[width=8px]{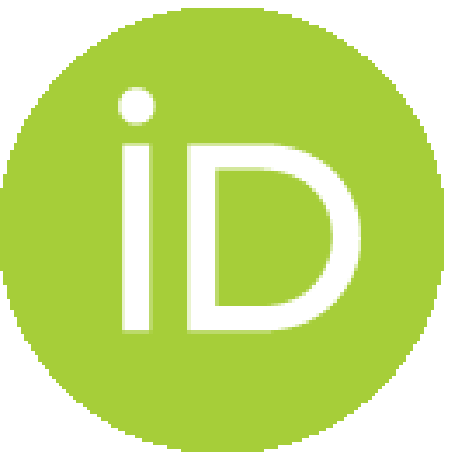}}}

\begin{document}

\title{21 cm power spectrum in interacting cubic Galileon model}

\author{Bikash R. Dinda \orcid{0000-0001-5432-667X}}
\email{bikashdinda.pdf@iiserkol.ac.in}
\email{bikashd18@gmail.com}
\affiliation{Department of Physical Sciences, Indian Institute of Science Education and Research Kolkata, India}
\affiliation{Department of Theoretical Physics, Tata Institute of Fundamental Research, Dr. Homi Bhabha Road, Navy Nagar, Colaba, Mumbai-400005, India}

\author{Md. Wali Hossain \orcid{0000-0001-6969-8716}}
\email{mhossain@jmi.ac.in}
\affiliation{Department of Physics, Jamia Millia Islamia, New Delhi, 110025, India}

\author{Anjan A. Sen}
\email{aasen@jmi.ac.in}
\affiliation{Centre For Theoretical Physics, Jamia Millia Islamia, New Delhi, 110025, India}

\begin{abstract}
We show the detectability of interacting and non-interacting cubic Galileon models from the $\Lambda$CDM model through the 21 cm power spectrum. We show that the interferometric observations like the upcoming SKA1-mid can detect both the interacting and the non-interacting cubic Galileon model from the $\Lambda$CDM model depending on the parameter values.
\end{abstract}

\maketitle
\date{\today}

\section{Introduction}
Recent cosmological observations \cite{Riess:1998cb,Perlmutter:1998np,Ade:2015xua} unveiled that our present Universe is going through an accelerated expanding phase. But to date, we do not have any proper theoretical explanation for this accelerated expansion. The simple possible explanation could be an exotic matter having negative pressure known as {\it dark energy} \cite{Copeland:2006wr,Linder:2008pp,Silvestri:2009hh,Sahni:1999gb}. The cosmological constant ($\Lambda$) is the simplest candidate for dark energy. It is also the most favored cosmological model observationally. However, it is plagued with the fine-tuning problem \citep{Martin:2012bt} and cosmic coincidence problem \citep{Zlatev:1998tr}. Apart from these well-known problems the concordance $\Lambda$CDM model is also in conflict with the string swampland conjectures \citep{Vafa:2005ui,Obied:2018sgi,Andriot:2018wzk} and the recent local measurement of the present value of the Hubble constant $H_0$ \citep{Riess:2019cxk,Wong:2019kwg,Pesce:2020xfe}. While the swampland conjecture rules out any stable de Sitter solution in string theory, the local measurement of $H_{0}$ points towards a discrepancy of 5$\sigma$ between $H_0$ constraint from the Planck observation of cosmic microwave background (CMB) with $\Lambda$CDM as the underlying model \citep{Planck:2018vyg} and the recent model independent local measurement of $H_0$ by Riess et al \citep{Riess:2019cxk}. 

To Look for alternative theories of gravity, one way is to modify the gravity at the large cosmological scale in such a way so that it becomes repulsive at large scales which gives rise to the accelerated expansion of the universe \cite{Clifton:2011jh,deRham:2014zqa,deRham:2012az,DeFelice:2010aj}. Dvali, Gabadadze, and Porrati (DGP) gave a scenario where a 4D Minkowsky brane is located on an infinitely large extra dimension and gravity is localized in the 4D Minkowsky brane \cite{Dvali:2000hr}. This model is known as the DGP model which gives rise to late time acceleration but its self-accelerating branch has a ghost \cite{Luty:2003vm,Nicolis:2004qq}. When we reduce the 5D DGP theory to 4D, at the decoupling limit, the theory gives rise to a Lagrangian of the form $(\nabla \phi)^2 \Box \phi$ \cite{Luty:2003vm}. This Lagrangian, in the Minkowski background, possesses the Galilean shift symmetry $\phi\to\phi+b_\mu x^\mu+c$, where $b_\mu$ and $c$ are the constants, and gives rise to second order equation of motion and hence free from ghost \cite{Luty:2003vm,Nicolis:2004qq,Nicolis:2008in}. Because of possessing the shift symmetry, the scalar field is dubbed as the "Galileon" \cite{Nicolis:2008in}. In the Minkowski background, we can construct two more Lagrangians containing higher derivatives and shift symmetry which gives rise to a second-order equation of motion. Along with a linear potential term and standard canonical kinetic term there exist five such terms which can possess the above-mentioned shift symmetry and give a second order equation of motion \cite{Nicolis:2008in}. All these five terms together form the Galileon Lagrangian \cite{Nicolis:2008in}. In the curved background, the non-linear completion of the Galileon Lagrangian includes non-minimal coupling to keep the equation of motion second order\cite{Deffayet:2009wt}. The non-minimal terms may cause the fifth force effect which can be evaded locally by implementing the Vainshtein mechanism \cite{Vainshtein:1972sx}. The Galileon models are the sub-classes of the more general scalar-tensor theory known as the Horndeski theory \cite{Horndeski:1974wa,Kobayashi:2011nu}. Late time cosmic acceleration is well studied in the Galileon theory  \cite{Chow:2009fm,Silva:2009km,Kobayashi:2010wa,Kobayashi:2009wr,Gannouji:2010au,DeFelice:2010gb,DeFelice:2010pv,Ali:2010gr,Mota:2010bs,Deffayet:2010qz,deRham:2010tw,deRham:2011by,Hossain:2012qm,Ali:2012cv}.

Horndeski theories are constraint by the detection of the event of binary neutron star merger GW170817, using both gravitational waves (GW) \cite{LIGOScientific:2017vwq} and its electromagnetic counterpart \cite{LIGOScientific:2017zic,LIGOScientific:2017ync}, which rules out a large class of Horndeski theories that predict the speed of GW propagation different from that of the speed of light \cite{Ezquiaga:2017ekz,Zumalacarregui:2020cjh}. The only higher derivative term that survives is $\sim G(\phi,X) \Box \phi$,where $X=-(1/2)(\nabla\phi)^2$ and $G(\phi,X)$ is a function of $\phi$ and $X$. When $G(\phi,X)\sim(\nabla \phi)^2$ it is cubic Galileon term. This cubic term along with the usual kinetic term and a potential term forms the Cubic Galileon model. Potentials other than the linear one break the shift symmetry but the equation of motion is still second order. These kinds of models are known as the Light Mass Galileon models \cite{Hossain:2012qm,Ali:2012cv}. Without the potential term, in the Cubic Galileon models, we can not have stable late time acceleration \cite{Gannouji:2010au}. This model has been studied extensively in the context of late time cosmology \cite{Chow:2009fm,Silva:2009km,Hossain:2012qm,Ali:2012cv,Brahma:2019kch,Bartolo:2013ws,Bellini:2013hea,Barreira:2013eea,Hossain:2017ica,Dinda:2017lpz}.

In literature, there are several dark energy and modified gravity models. Good dark energy or a modified gravity model should be consistent with different cosmological observations. Hence it is important to study the cubic Galileon model in the context of cosmological observations. In literature, such efforts involving the Galileon model have been done earlier, for example, in the context of type Ia supernova observations \citep{Brahma:2020eqd}, cosmic microwave background, and baryonic acoustic oscillations observations \citep{Renk:2017rzu} etc. In the same spirit, the 21 cm cosmological observations like using the upcoming SKA telescope (square kilometer array) will be promising to detect dark energy and modification of gravity models. For this purpose, the post-reionization epoch (redshift $<6$) is particularly important to constrain the dynamics of dark energy or the dynamics of cosmological geometry in the modified gravity models. In the post-reionization epoch, the Universe is assumed to be highly ionized but neutral hydrogen atoms (HI) are still present and these HI are the biased tracers of the matter in the Universe. Thus the HI energy density tracks the matter-energy density in the Universe and the HI power spectrum is related to the matter power spectrum. The power spectrum of the intensity mapping of the large-scale HI distribution (commonly called the 21 cm power spectrum) is thus one kind of measurement of the large-scale structure formation \citep{Wyithe:2007gz,Loeb:2008hg}.

This paper aims to study the effect of the cubic Galileon model in the 21 cm power spectrum and to check the detectability of the deviation due to this model from the $\Lambda$CDM in the context of the upcoming SKA1-MID telescope specification. The upcoming SKA1-MID telescope is specifically proposed designed to study the structure formation in the low redshift regions ($<3$) which is useful to constrain dark energy and modified gravity model parameters \citep{SKA:2018ckk}. According to the updated proposed design of SKA1-MID, it will have a total of 197 antennas. Among these 64 antennas are from MeerKAT and 133 antennas are originally from SKA1-MID \citep{SKA:2018ckk}. SKA1-MID is proposed to be detecting the redshifted 21 cm line signal at the redshift range from 0.5 to 3 which will have good accuracy (competitive or even with better accuracy than other observations related to galaxy clustering etc.) to test the dynamics of the dark energy or modified gravity \citep{SKA:2018ckk,2019arXiv191212699B}.

The paper is organized as follows: in Sec.~\ref{sec-background} we describe the background dynamics of the Universe in presence of the interacting cubic Galileon field; in Sec.~\ref{sec-perturbation} we describe the evolution of the matter inhomogeneity and the corresponding matter power spectrum; in Sec.~\ref{sec-21cmps} we describe the 21 cm power spectrum; in Sec.~\ref{sec-ska1mid} we show the detectability of interacting cubic Galileon model in the SKA1-MID telescope specifications; finally in Sec.~\ref{sec-conclusion} we present our conclusion.

\section{Background evolution}
\label{sec-background}
We consider the following action in the Einstein frame with a potential $V(\phi)$ \cite{Ali:2012cv}
\begin{eqnarray}
\S=\int {\rm d}^4x\sqrt{-{\rm g}}\Big [\frac{\Mpl^2}{2} R-\frac{1}{2}(\nabla \phi)^2\Bigl(1+\frac{\al}{M^3}\Box \phi\Bigr) - V(\phi) \Big]+ \S_\m\Bigl[\Psi_\m;{\rm e}^{2 \beta \phi/M_{\rm pl}} {\rm g}_{\mu\nu}\Bigr] \, ,
\label{eq:action}
\end{eqnarray}
where the scalar field is non-minimally coupled with gravity in the Jordan frame with $\beta$ as the coupling constant and ${\rm e}^{2 \beta \phi/M_{\rm pl}}$ is the conformal factor that relates the Jordan and Einstein frame metric tensors. In Eq~\eqref{eq:action} $M$ is a constant of mass dimension one, $\Mpl=1/\sqrt{8\pi G}$ is the reduced Planck mass, $\al$ is a dimensionless constant and $\S_\m$ corresponds to the matter action with $\Psi_\m$'s as the matter fields. Action~\eqref{eq:action} can be realized as a sub-class of Horndeski theories \citep{Horndeski:1974wa,Kobayashi:2011nu} and one can recover the usual quintessence models on taking $\alpha\rightarrow0$. $V(\phi)$ is the potential of the cubic Galileon field. Throughout the paper, we have considered only the linear potential, because Galileon shift symmetry is preserved only in linear potential.

In flat Friedmann--Lema\^itre--Robertson--Walker (FLRW) metric, given by $ds^{2} = - dt^{2} + a^{2} (t) d\vec{r}.d\vec{r}$, where $t$ is the cosmic time, $\vec{r}$ is the comoving coordinate vector and $a$ is the cosmic scale factor, the background cosmological equations can be obtained by varying action \eqref{eq:action} with respect to the metric tensor ${\rm g_{\mu\nu}}$ \citep{Ali:2012cv,Hossain:2012qm}

\begin{eqnarray}
3M_{\rm pl}^2H^2 &=& \bar{\rho}_m+\frac{\dot{\phi}^2}{2}\Bigl(1-6 \frac{\alpha}{M^3} H\dot{\phi}\Bigr)+V{(\phi)},
\label{eq:first_Friedmann} \\
M_{\rm pl}^2(2\dot H + 3H^2) &=& -\frac{\dot{\phi}^2}{2}\Bigl(1+2 \frac{\alpha}{M^3} \ddot{\phi}\Bigr)+V(\phi),
\label{eq:second_Friedmann}
\end{eqnarray}
 
\noindent
where over-dot is the derivative with respect to the cosmic time $t$, $H$ is the Hubble parameter and $ \bar{\rho}_m $ is the background matter-energy density. The background equation of motion for the Galileon field $ \phi $ is given by \citep{Ali:2012cv}

\begin{equation}
\ddot{\phi} + 3H\dot{\phi}-3 \frac{\alpha}{M^3} \dot{\phi}\Bigl(3H^2\dot{\phi}+\dot{H}\dot{\phi}+2H\ddot{\phi}\Bigr)+ V_{\phi}= - \frac{\beta}{M_{Pl}} \bar{\rho}_m ,
\label{eq:E-L_eq}
\end{equation}

\noindent
where subscript $\phi$ is the derivative with respect to the field $\phi$. The continuity equations for matter and scalar field are given by
\begin{eqnarray}
\dot{ \bar{\rho}}_m + 3 H \bar{\rho}_m &=& \frac{\beta}{M_{Pl}} \dot{\phi} \bar{\rho}_m \, ,\\
\dot{ \bar{\rho}}_{\phi} + 3 H (1+w_{\phi}) \bar{\rho}_{\phi} &=& - \frac{\beta}{M_{Pl}} \dot{\phi} \bar{\rho}_m \, ,
\end{eqnarray}

\noindent
where $w_{\phi}$ is the equation of the state of the scalar field. 

To study the background evolution we rewrite the above differential equations as an autonomous system of equations. To do this we define the following dimensionless variables \citep{Ali:2012cv,Hossain:2012qm}

\begin{eqnarray}
x &=& \frac{ \dot{\phi} }{\sqrt{6} H M_{Pl}} = \frac{\Big{(} \dfrac{d \phi}{d N} \Big{)}}{\sqrt{6} M_{Pl}}\, , 
\label{eq:x}\\
y &=& \frac{\sqrt{V}}{\sqrt{3} H M_{Pl}}\, , 
\label{eq:y}\\
\epsilon &=& -6 \frac{\alpha}{M^3} H \dot{\phi} = -6 \frac{\alpha}{M^3} H^{2} \Big{(} \dfrac{d \phi}{d N} \Big{)}\, , 
\label{eq:ep}\\
\lambda &=& - M_{Pl} \frac{V_{\phi}}{V}, 
\label{eq:lam}\\
\Gamma &=& \frac{V_{\phi \phi}V}{V_{\phi}^{2}}\, ,
\label{eq:gam}
\end{eqnarray}
where $N= \ln a$. Using the above-mentioned dimensionless variables we can form the following autonomous system \citep{Ali:2012cv}

\begin{align}
\label{eq:auto_diff}
\frac{{\rm d}x}{{\rm d}N}&=x\Bigl(\frac{\ddot{\phi}}{H\dot{\phi}}-\frac{\dot H}{H^2}\Bigr), \nonumber\\
\frac{{\rm d}y}{{\rm d}N}&=-y \Bigl(\sqrt{\frac{3}{2}}\lambda x+\frac{\dot H}{H^2}\Bigr), \nonumber\\
\frac{{\rm d}\epsilon}{{\rm d}N}&=\epsilon \Bigl(\frac{\ddot{\phi}}{H\dot{\phi}}+\frac{\dot H}{H^2}\Bigr), \nonumber\\
\frac{{\rm d}\lambda}{{\rm d}N}&=\sqrt{6}x\lambda^2(1-\Gamma),
\end{align}
where
\begin{align}
\frac{\dot H}{H^2}&=\frac{6(1+\epsilon)(y^2-1)-3x^2(2+4\epsilon+\epsilon^2)}{4+4\epsilon+x^2\epsilon^2} +\frac{\sqrt{6}x\epsilon (y^2\lambda -\beta \Omega_m)}{4+4\epsilon+x^2\epsilon^2}\\
\frac{\ddot{\phi}}{H\dot{\phi}}&=\frac{3x^3\epsilon-3x\Bigl(4+\epsilon (1+y^2)\Bigr)+2\sqrt{6}(y^2\lambda-\beta\Omega_m)}{x(4+4\epsilon+x^2\epsilon^2)}
\end{align}

\noindent
From Eq.~\eqref{eq:first_Friedmann} we have the constraint equation $\Omega_\m+\Omega_\phi=1$, where $\Omega_\m$ is the matter density parameter and $\Omega_{\phi} = x^2 (\epsilon +1)+y^2$ is the scalar field density parameter. The scalar field ($w_\phi$) and the effective ($w_{\rm eff}$) equation of states are given respectively

\begin{eqnarray}
w_{\phi} &=& \frac{x \left(3 x (\epsilon  (\epsilon +8)+4)-2 \sqrt{6} \beta  \epsilon  \left(x^2 (\epsilon +1)-1\right)\right)-2 y^2 \left(\epsilon  \left(\sqrt{6} x (\beta +\lambda )+6\right)+6\right)}{3 \left(\epsilon  \left(x^2 \epsilon +4\right)+4\right) \left(x^2 (\epsilon +1)+y^2\right)},
\label{eq:wphi} \\
w_{ \text{eff} } &=& \frac{x \left(3 x (\epsilon  (\epsilon +8)+4)-2 \sqrt{6} \beta  \epsilon  \left(x^2 (\epsilon +1)-1\right)\right)-2 y^2 \left(\epsilon  \left(\sqrt{6} x (\beta +\lambda )+6\right)+6\right)}{3 \left(\epsilon  \left(x^2 \epsilon +4\right)+4\right)}.
\label{eq:weff}
\end{eqnarray}

To solve the above system of differential equations (Eq.~\eqref{eq:auto_diff}), we need initial conditions for $x$, $y$, $\epsilon$ and $\lambda$. We denote these by $x_i$, $y_i$, $\epsilon_i$ and $\lambda_i$ respectively. In literature, scalar field models are mainly two types depending on the attractor or thawing behavior. Here, we consider the thawing type of behavior for the Galileon field and accordingly choose the initial conditions. In the thawing kind of behavior, the equation of state of the scalar field is very close to $-1$ at initial times, and at late times it thaws away from $-1$. From Eq.~\eqref{eq:wphi}, we can see that $w_{\phi} \approx -1$ when $x \ll y$. We want this behavior at initial times so we fix $x_i = 10^{-3} y_i$ throughout. We fix the initial conditions at redshift, $z=1000$. Next, we fix the $y_i$ parameter for which $\Omega_{m0}=0.3111$. This value corresponds to the Planck 2018 results. Here $\Omega_{m0}$ is the present value of the matter-energy density parameter. So, we are left with three parameters $\epsilon_i$, $\lambda_i$, and the $\beta$ parameters. $\epsilon_i$ represents the deviation from the quintessence model, $\lambda_i$ represents the initial slope of the potential and $\beta$ represents the interaction between matter and the Galileon field.

\begin{figure}[tbp]
\centering
\includegraphics[width=.45\textwidth]{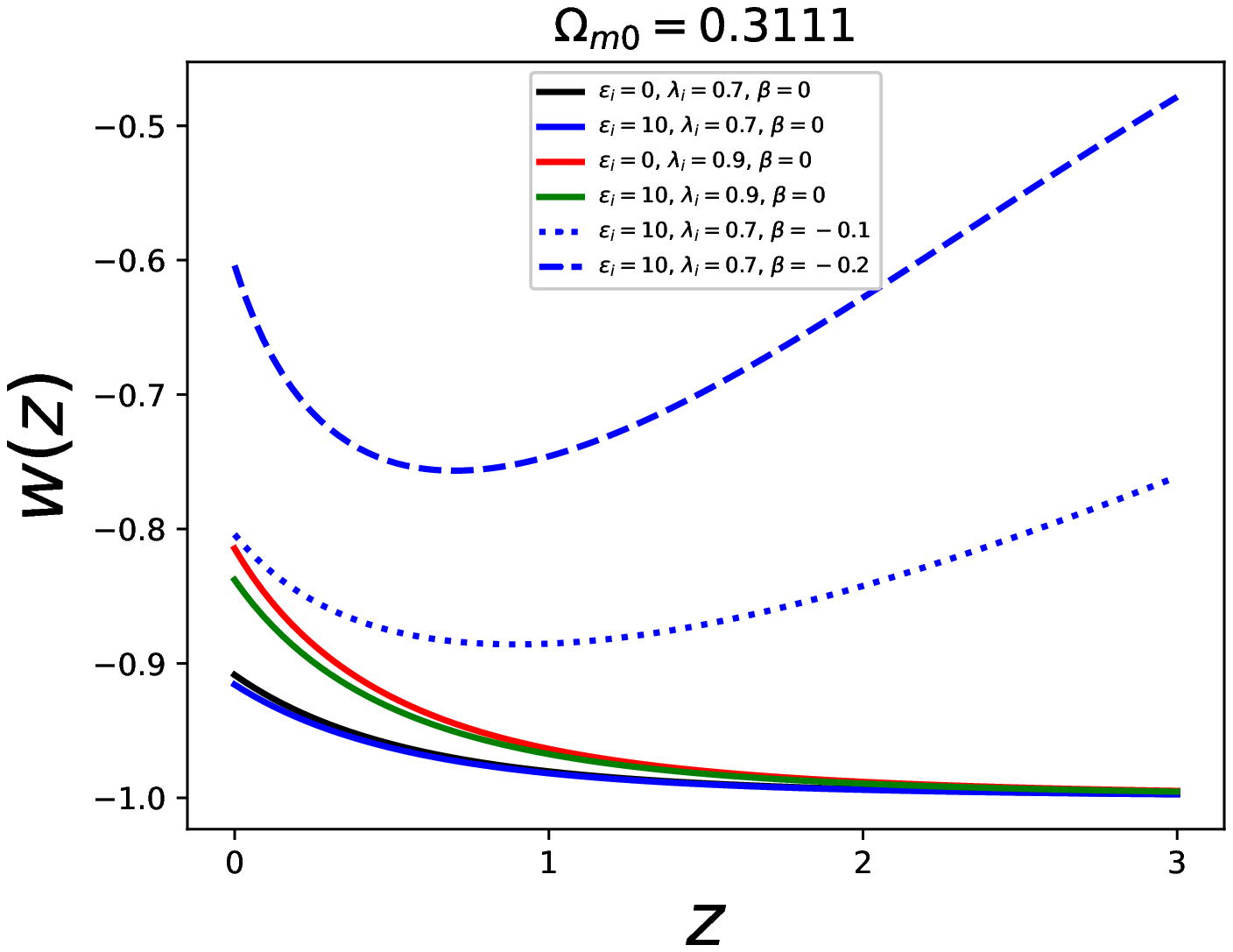}
\includegraphics[width=.45\textwidth]{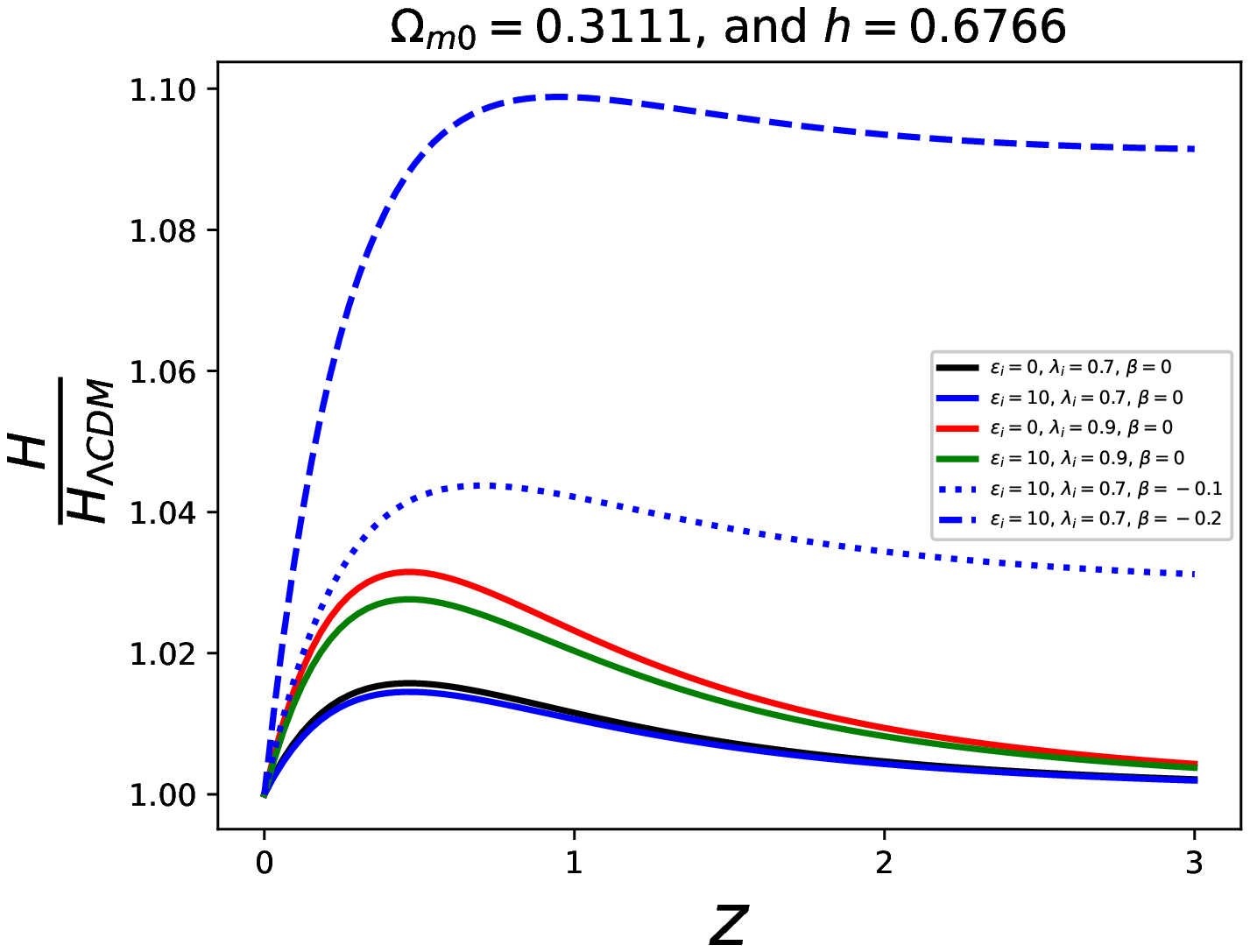}
\caption{\label{fig:wphi_H} Background evolution of the cubic Galileon field. Left panel: equation of state of the cubic Galileon field. Right panel: comparison of the Hubble parameter with the $\Lambda$CDM one.}
\end{figure}

With the above-mentioned initial conditions, we show the background evolutions in the interacting cubic Galileon model, by plotting the equation of state and the Hubble parameter in Figure~\ref{fig:wphi_H}. From the left panel, we see that the higher the $\epsilon_i$ value lesser the deviation from $\Lambda$CDM. Higher the $\lambda_i$ value larger the deviation from $\Lambda$CDM. More negative the $\beta$ value larger the deviation from the $\Lambda$CDM. We see the same behavior in the deviation of the Hubble parameter in the right panel.

\section{Evolution of perturbations}
\label{sec-perturbation}

In this work, we are working in the sub-Huuble limit. In this limit, we can use the quasi-static approximation i.e we can consider Newtonian perturbation. With the Newtonian perturbation and in the linear theory, the evolution equation of the growth function ($D_{+}$) for the cubic Galileon model is given by \citep{gal1,gal6}

\begin{equation}
\dfrac{d^{2} D_{+}}{d N^{2}} + \left[ - \frac{1}{2} 3 (w_{\text{eff}}+1)+\sqrt{6} \beta  x+2 \right] \dfrac{d D_{+}}{d N} - \frac{3}{2} \Omega_{m} \frac{G_{ \text{eff} }}{G} D_{+} = 0 ,
\label{eq:Dplus}
\end{equation}

\noindent
where the growth function is defined as $ \delta_{m} (z) = D_{+} (z) \delta_{m}^{i} $. $\delta_{m}$ is the matter inhomogeneity and $\delta_{m}^{i}$ its initial value. $G$ is the Newtonian gravitational constant. In the interacting cubic Galileon model, the effective gravitational constant ($G_{ \text{eff} }$) is given by \citep{Ali:2012cv}

\begin{equation}
\frac{G_{ \text{eff} }}{G} = \frac{x \left(3 x (\epsilon  (\epsilon +8)+4)-2 \sqrt{6} \beta  \epsilon  \left(x^2 (\epsilon +1)-1\right)\right)-2 y^2 \left(\epsilon  \left(\sqrt{6} x (\beta +\lambda )+6\right)+6\right)}{3 \left(\epsilon  \left(x^2 \epsilon +4\right)+4\right)} .
\end{equation}

\noindent
To solve the differential equation of the growth function, we consider the fact that at an initial time, at matter dominated epoch (here at $z=1000$), $ D_{+} \propto a $. This corresponds to $ D_{+}|_{i} = \frac{1}{1+1100} = \frac{d D_{+}}{d N} \big{|}_{i} $.


Once we have the solution for the growth function, we can compute the linear matter power spectrum $ P_{m} $ given by

\begin{equation}
P_{m}(k,z) = A k^{n_{s}} T^{2} (k) \frac{D_{+}^{2}(z)}{D_{+}^{2}(z=0)},
\label{eq:Pm}
\end{equation}

where $k$ corresponds to the amplitude of the wave vector. $n_{s}$ is the scalar spectral index for the primordial power spectrum. $ T(k) $ is the transfer function. For our calculations, we consider the $T(k)$ given by Eisenstein and Hu \citep{eisenhu}. $A$ is the normalization constant corresponds to the usual $ \sigma_{8} $ normalization. Here, we consider Eisenstein-Hu transfer function and for the $\sigma_{8}$ normalisation, we fix $\Omega_{m}^{(0)}=0.3111$, $h=0.6766$, $\Omega_{b}^{(0)}=0.049$, $n_{s}=0.9665$ and $\sigma_{8}=0.8102$. These values are best-fit values according to the Planck 2018 results. $h$ is defined as $H_0 = 100 h Km S^{-1} Mpc^{-1}$ with $H_0$ being the present value  of the Hubble parameter. $\Omega_{b}^{(0)}$ is the present value of the baryonic matter energy density parameter.

\begin{figure}[tbp]
\centering
\includegraphics[width=.45\textwidth]{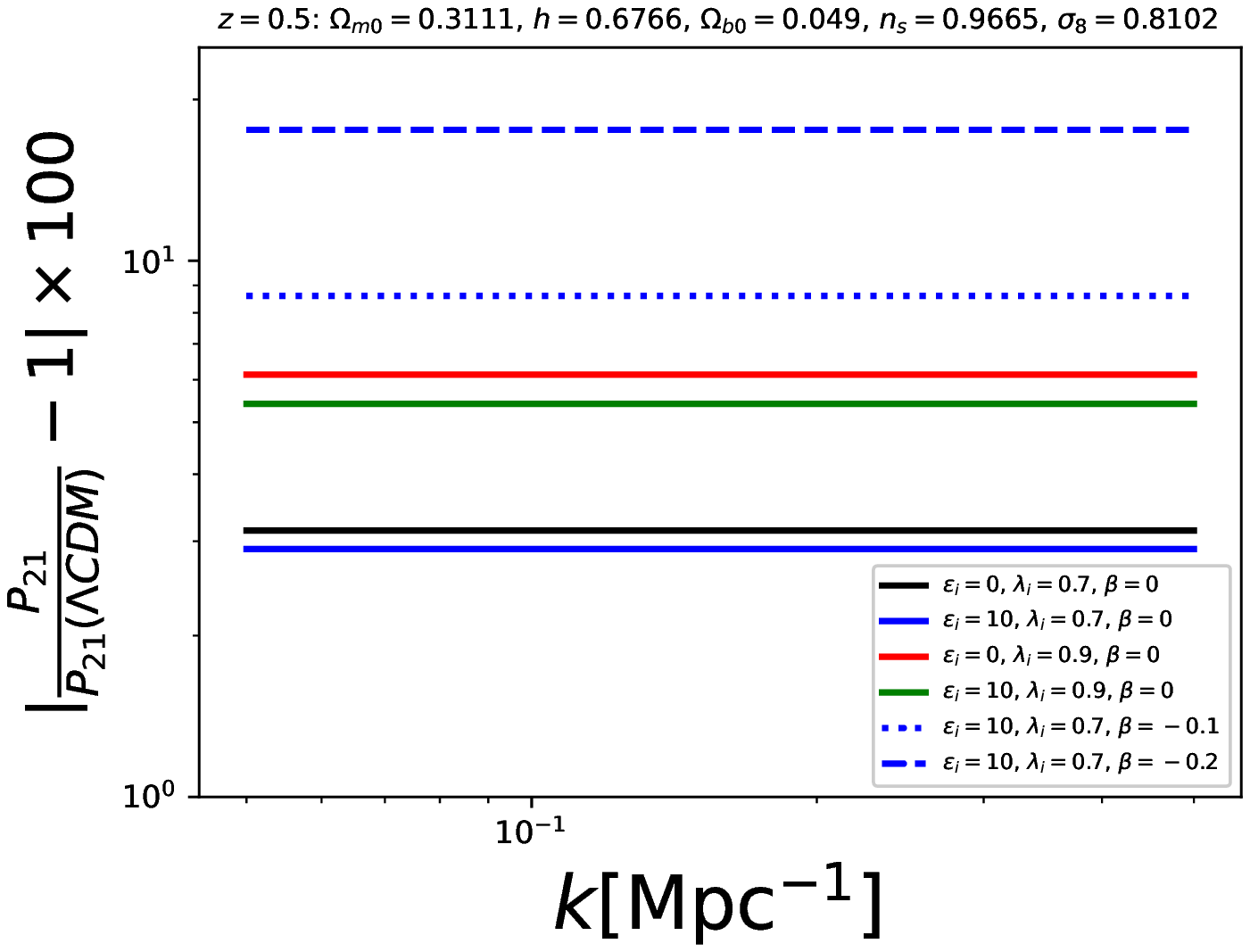}
\includegraphics[width=.45\textwidth]{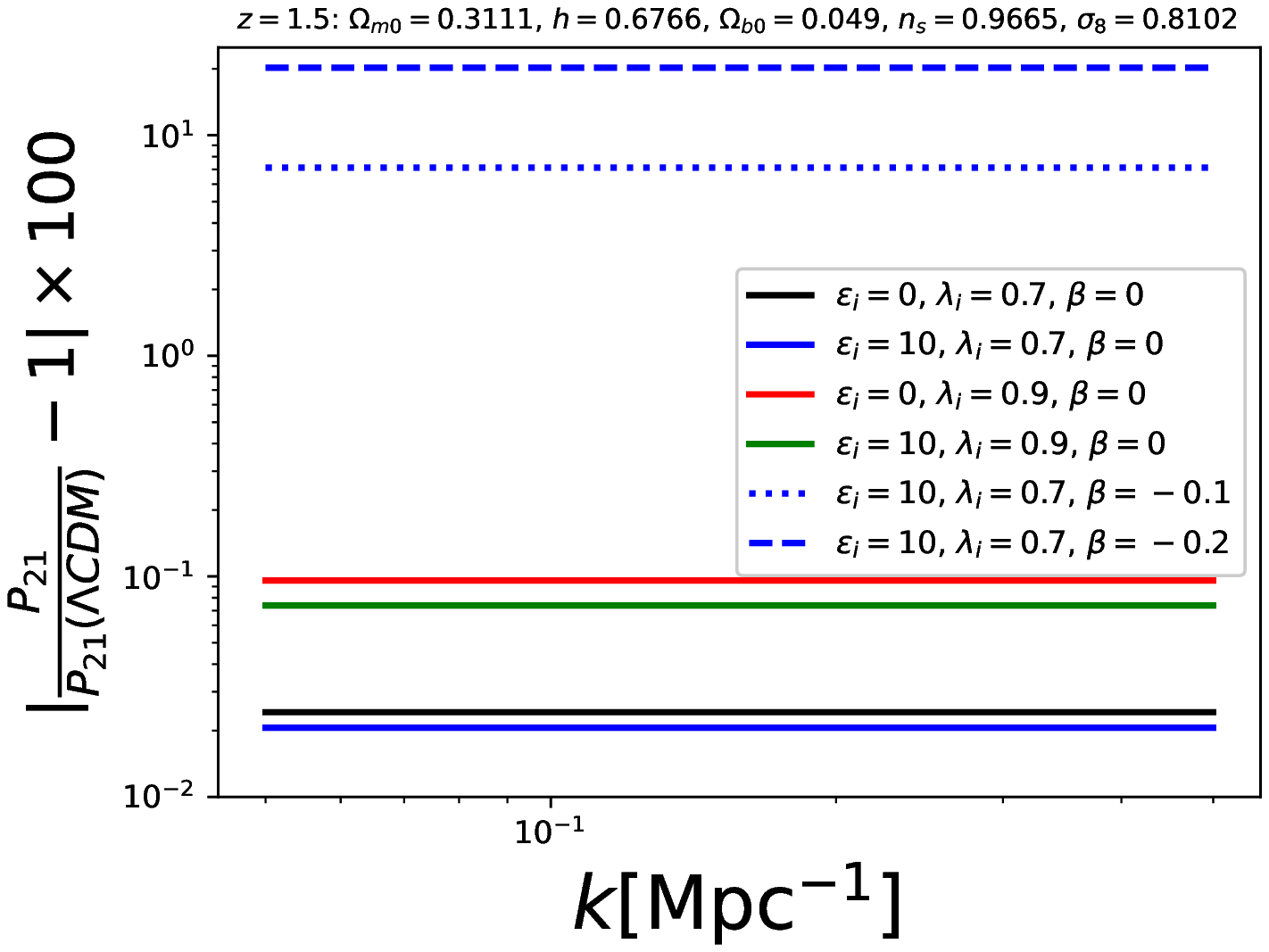}\\
\includegraphics[width=.45\textwidth]{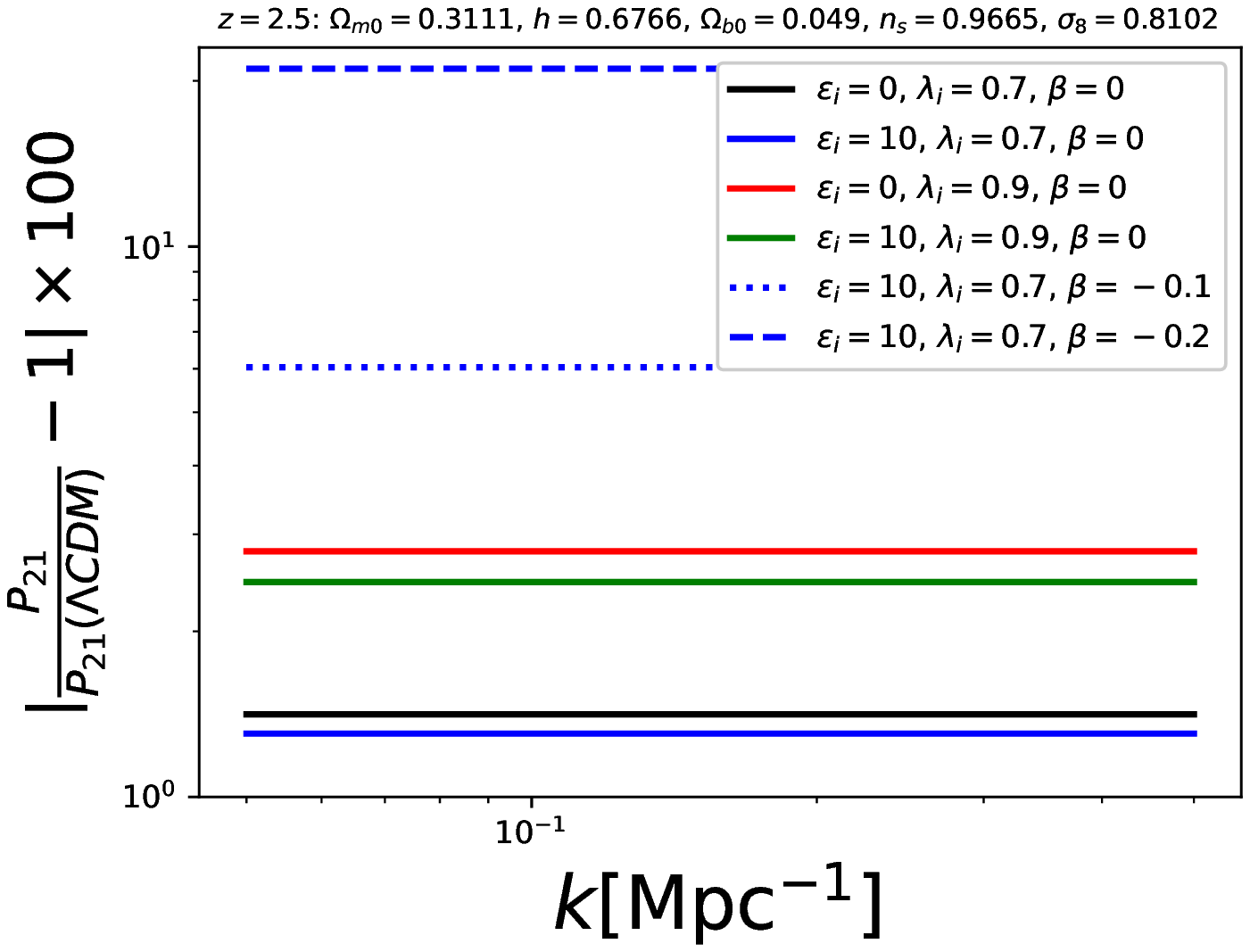}
\caption{\label{fig:P21cmp} Percentage deviation in the angle averaged 21 cm power spectrum for the cubic Galileon model from $ \Lambda $CDM model.}
\end{figure}

\begin{figure}[tbp]
\centering
\includegraphics[width=.45\textwidth]{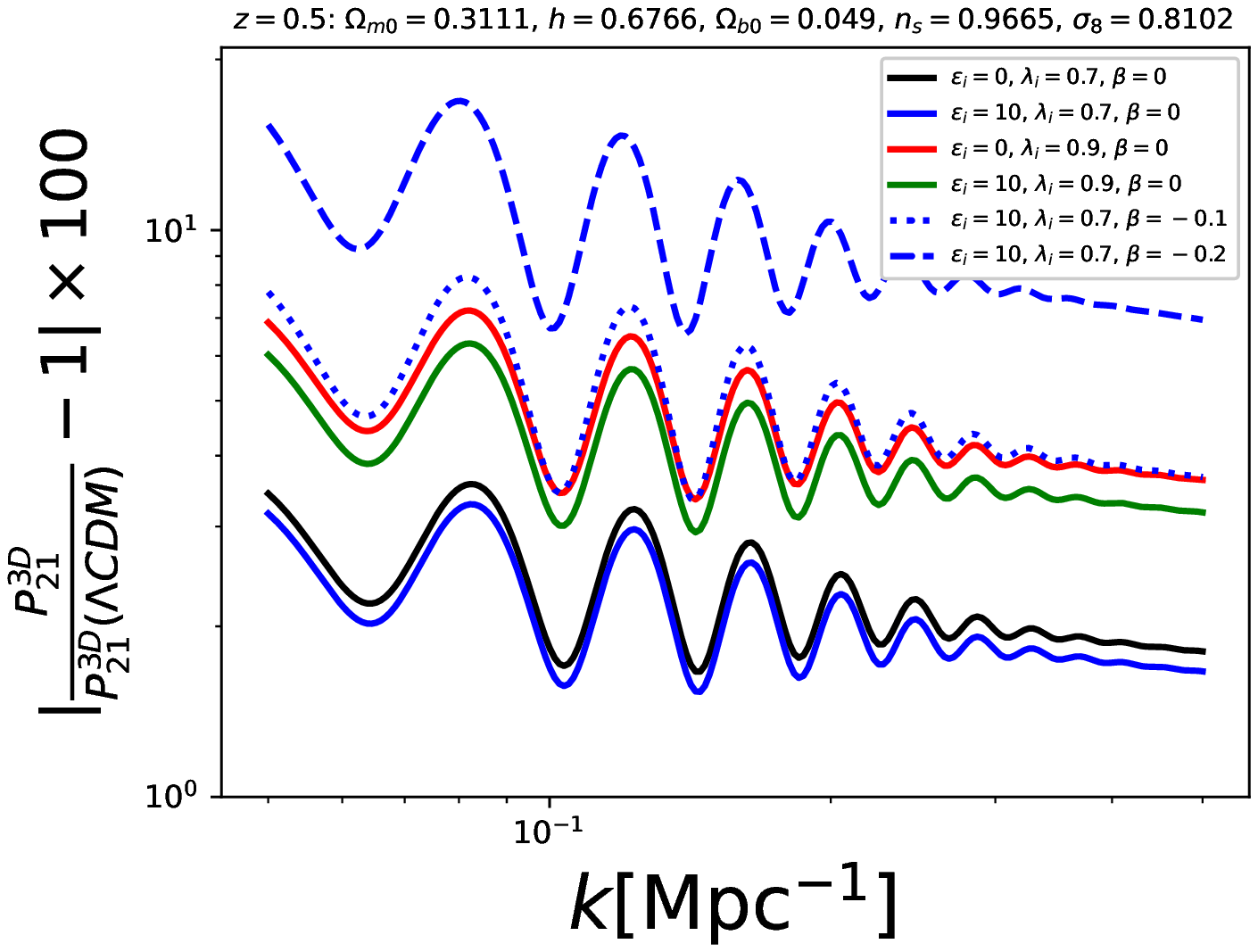}
\includegraphics[width=.45\textwidth]{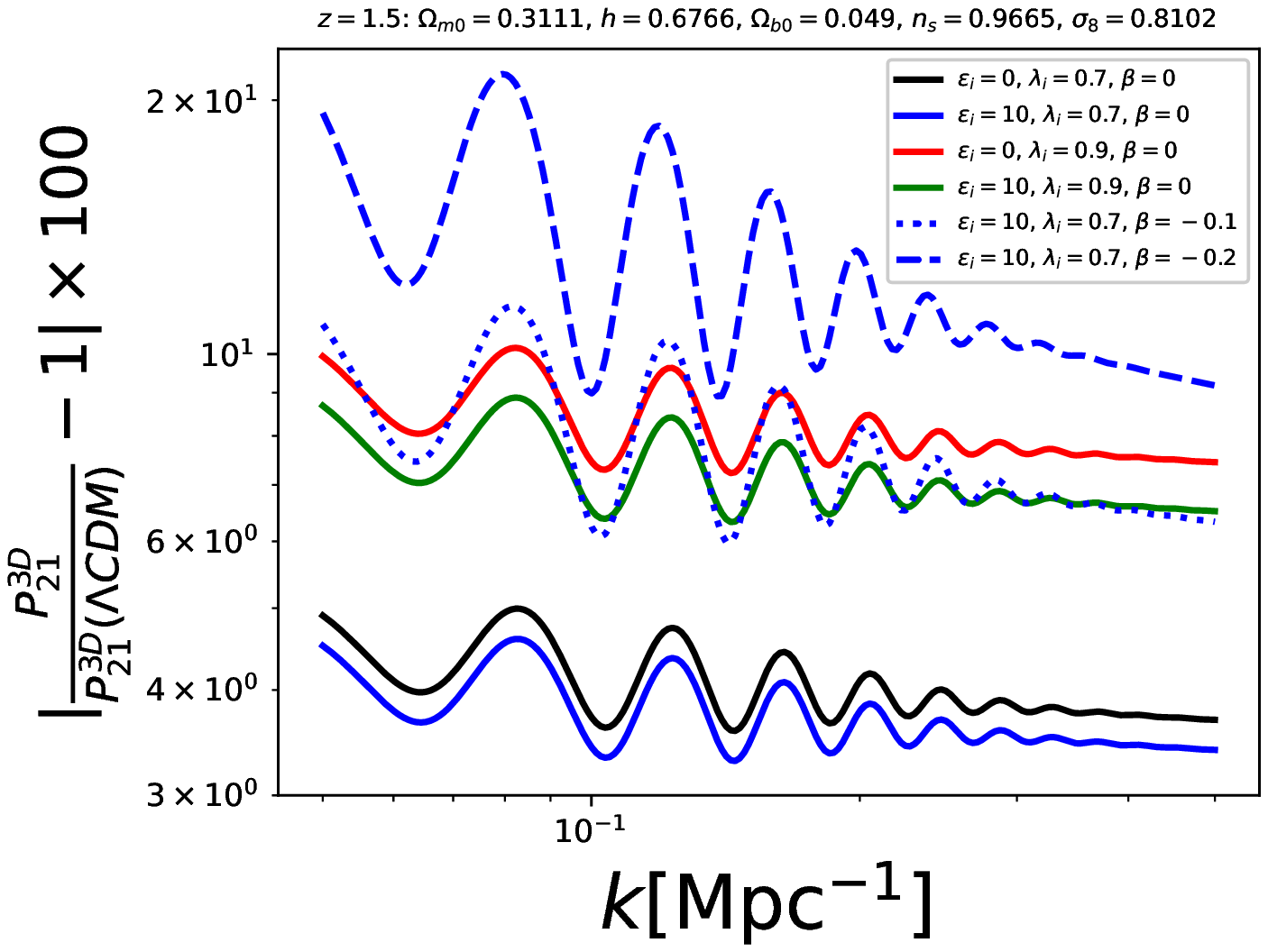}\\
\includegraphics[width=.45\textwidth]{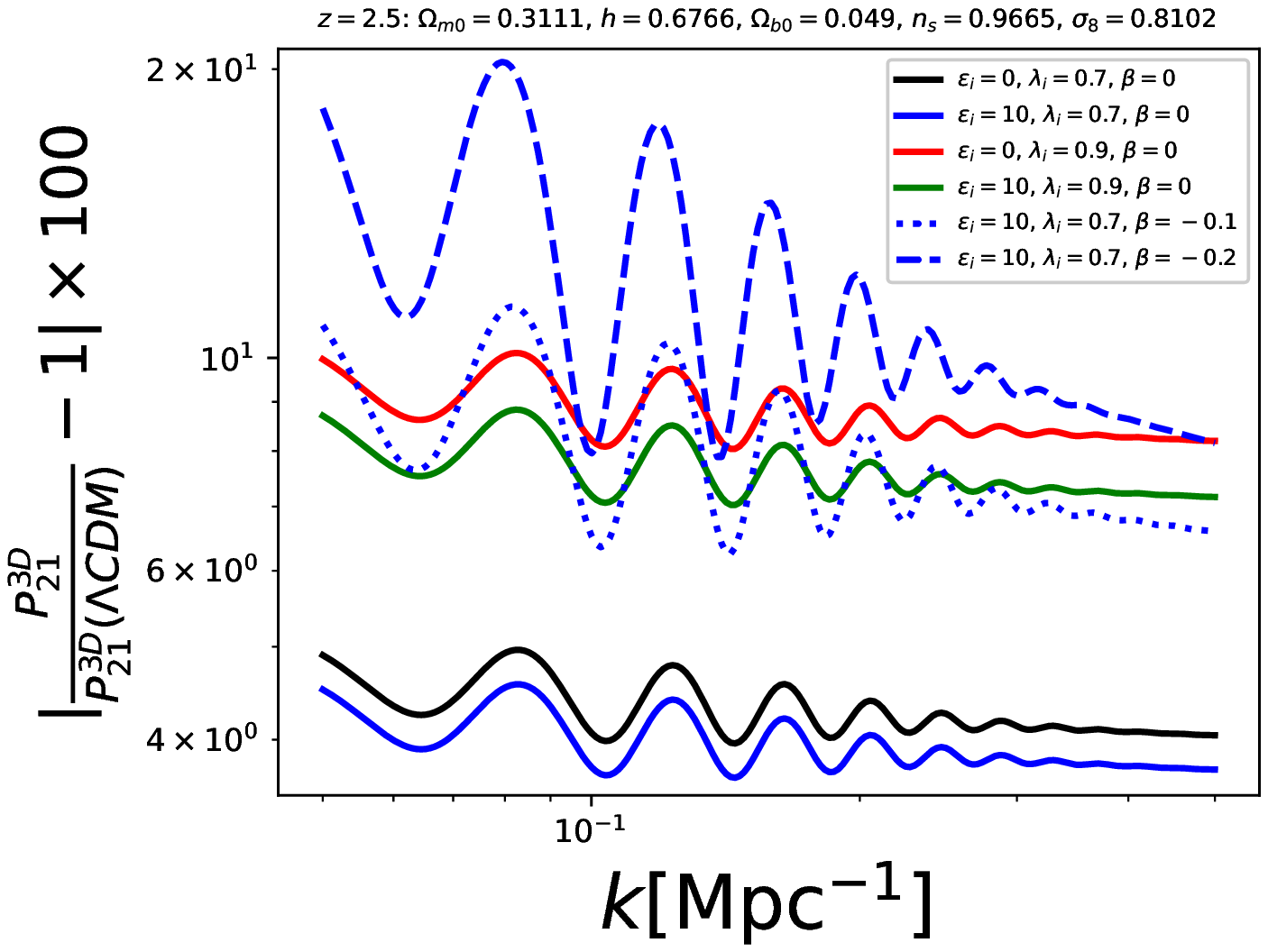}
\caption{\label{fig:ps21D3Lcdm} Percentage deviation in the angle averaged 3D 21 cm power spectrum for the cubic Galileon model from the $\Lambda$CDM model (here fiducial model).}
\end{figure}

\section{21 cm power spectrum}
\label{sec-21cmps}
The matter distribution is not directly related to the observation. However, in this regard 2-point correlation in excess brightness temperature is useful. So, we study the detectability of the cubic Galileon model over the $\Lambda$CDM model using the 21 cm power spectrum. The 21 cm power spectrum, $P_{21}$ of the excess brightness temperature field is given by \citep{PS_21_1,PS_21_2,PS_21_3,PS_21_5,PS_21_6,PS_21_7,PS_21_8,PS_21_10}

\begin{equation}
P_{21} (k,z,\mu) = C_{T}^{2} (1+\beta_{T} \mu^{2})^{2} P_{m}(k,z),
\label{eq:P212D}
\end{equation}

\noindent
where $\mu=\hat{n}.\hat{k}=\cos{\theta}$, where $\hat{n}$ is the line of sight (LOS) unit direction and $\theta$ is the angle between LOS and the wave vector. $\beta_T$ is defined as

\begin{equation}
\beta_T = \frac{f}{b},
\label{eq:betaT}
\end{equation}

\noindent
where $f$ is the growth factor and it is defined as $f=\frac{d \ln{D_{+}}}{d \ln{a}}$. $b$ is the linear bias that connects the HI distribution to the matter distribution. Throughout this paper, we consider the linear bias i.e. $b=1$. $C_{T}$ is the mean HI excess brightness temperature given by \citep{PS_21_1,PS_21_2}

\begin{equation}
C_{T} (z) = b \bar{x}_{HI} \bar{T} (z),
\label{eq:CT}
\end{equation}

\noindent
where $\bar{x}_{HI}$ is the neutral hydrogen fraction. $\bar{T}$ is given by \citep{PS_21_1,PS_21_2}

\begin{equation}
\bar{T} (z) = 4.0 mK (1+z)^{2} \Big{(} \frac{\Omega_{b0} h^{2}}{0.02} \Big{)} \Big{(} \frac{0.7}{h} \Big{)} \frac{H_{0}}{H(z)}.
\label{eq:Tbar}
\end{equation}

$\mu$ averaged 21 cm power spectrum is computed as

\begin{eqnarray}
P_{21}(k,z) = \int_{0}^{1} d\mu \hspace{0.2 cm} P_{21}(k,z,\mu).
\label{eq:P21avg}
\end{eqnarray}

\noindent
Note that we keep the same notation, $P_{21}$ for the $\mu$ averaged 21 cm power spectrum. We show the deviations in $P_{21}(k,z)$ in Figure~\ref{fig:P21cmp} for the cubic Galileon model from $\Lambda$CDM with the same combinations of parameter values as in Figure~\ref{fig:wphi_H}. We see that the deviations are up to 2 to 20$\%$ depending on the parameters.

In radio interferometric observations like in square kilometer array (SKA) observation, the observables are not measured in $k$, but rather in the baseline distribution, $U$. The conversion from $k$ to $U$ is fiducial cosmological model dependent. If this fiducial model is different from the actual model, we need to consider the correction to the 21 cm power spectrum, defined in Eq.~\eqref{eq:P212D}. Throughout this paper, we consider $\Lambda$CDM as the fiducial model. So, for the cubic Galileon model, the observed 21 cm power spectrum would be \citep{Bull,P21_3D_2,P21_3D_5}

\begin{equation}
P_{21}^{3D}(k,z,\mu) = \frac{1}{\alpha_{||} \alpha_{\perp}^{2}} C_{T}^{2} \Big{[} 1+\beta_{T} \frac{\mu^{2} / F^{2}}{1+(F^{-2}-1) \mu^{2}} \Big{]}^{2} P_{m} \Big{(} \frac{k}{\alpha_{\perp}} \sqrt{1+(F^{-2}-1) \mu^{2}}, z \Big{)},
\label{eq:P213D}
\end{equation}

\noindent
where $ \alpha_{||} = H_{\rm fd} / H $, $ \alpha_{\perp} = r / r_{\rm fd} $ and $ F = \alpha_{||} / \alpha_{\perp} $. The subscript 'fd' corresponds to the fiducial model. Here $r$ is the line of sight comoving distance. The above corrected 21 cm power spectrum is sometimes referred to as the 3D 21 cm power spectrum and that is why we denote this by putting a superscript '3D'. Now we compute the angle averaged (or $\mu$ averaged) 21 cm power spectrum as

\begin{eqnarray}
P_{21}^{3D}(k,z) = \int_{0}^{1} d\mu \hspace{0.2 cm} P_{21}^{3D}(k,z,\mu).
\label{eq:P213Davg}
\end{eqnarray}

\noindent
Note that, we use the same notation for averaged 21 cm power spectrum.

In Figure~\ref{fig:ps21D3Lcdm}, we show the deviation in the angle averaged observed (3D) 21 cm power spectrum for the cubic Galileon model from the $\Lambda$CDM model for the same parameter values as in Figure~\ref{fig:P21cmp}. In the 3D 21 cm power, the deviations are similar. Another important thing to notice here is that there is $k$ dependence in the deviations in the 3D power spectrum unlike in the normal power spectrum. This is because the fiducial model and the actual model are different here.

\begin{figure}[tbp]
\centering
\includegraphics[width=.45\textwidth]{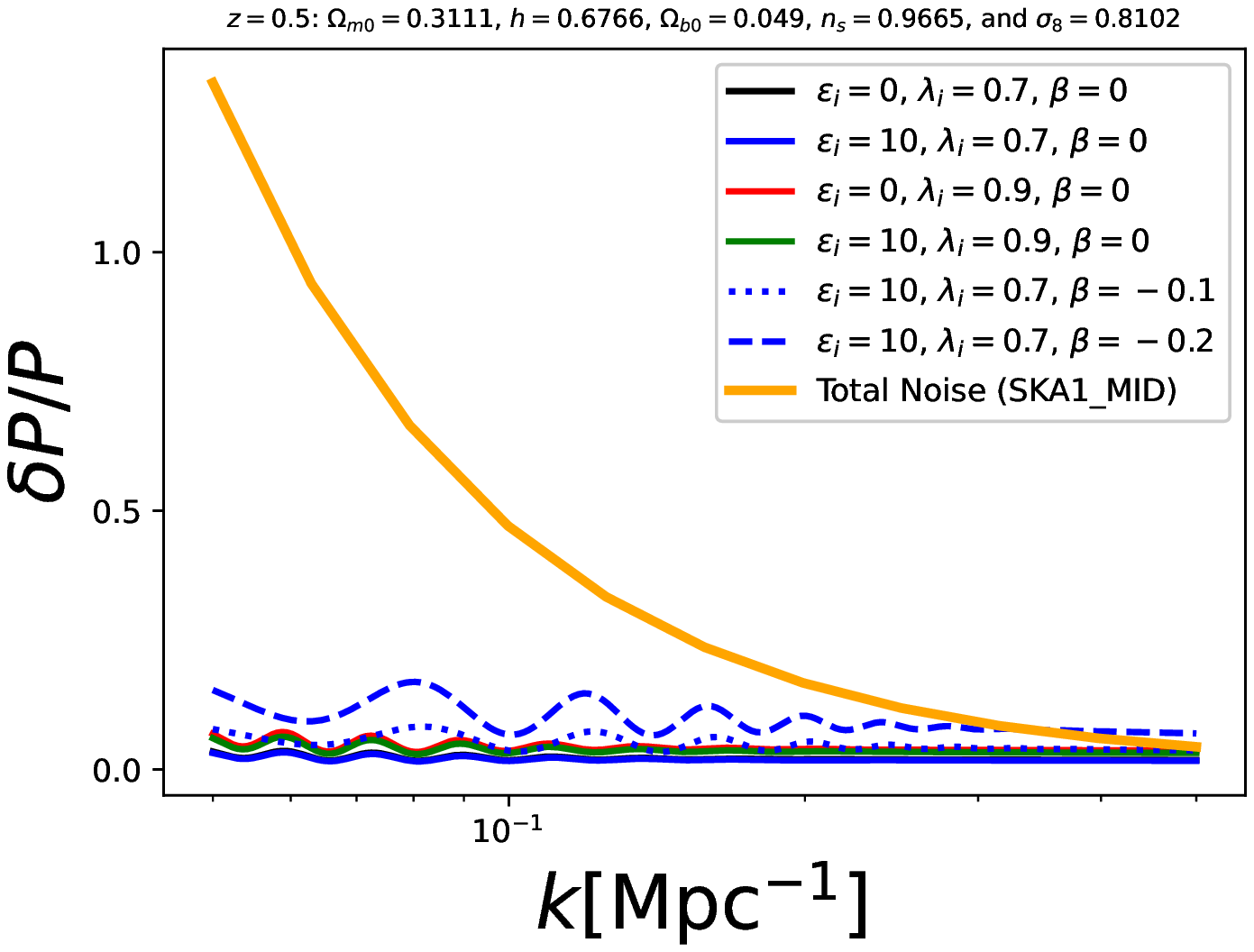}
\includegraphics[width=.45\textwidth]{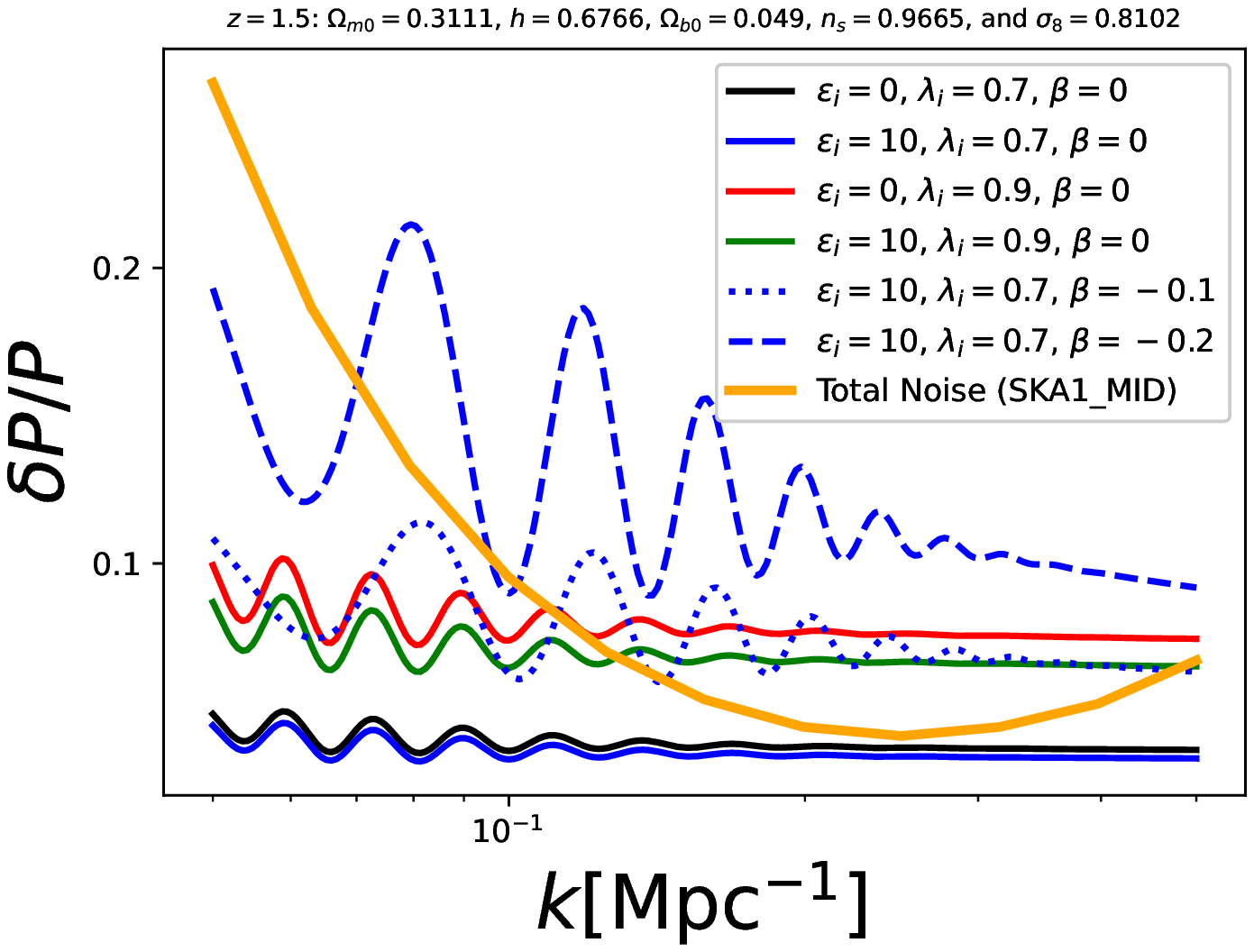}\\
\includegraphics[width=.45\textwidth]{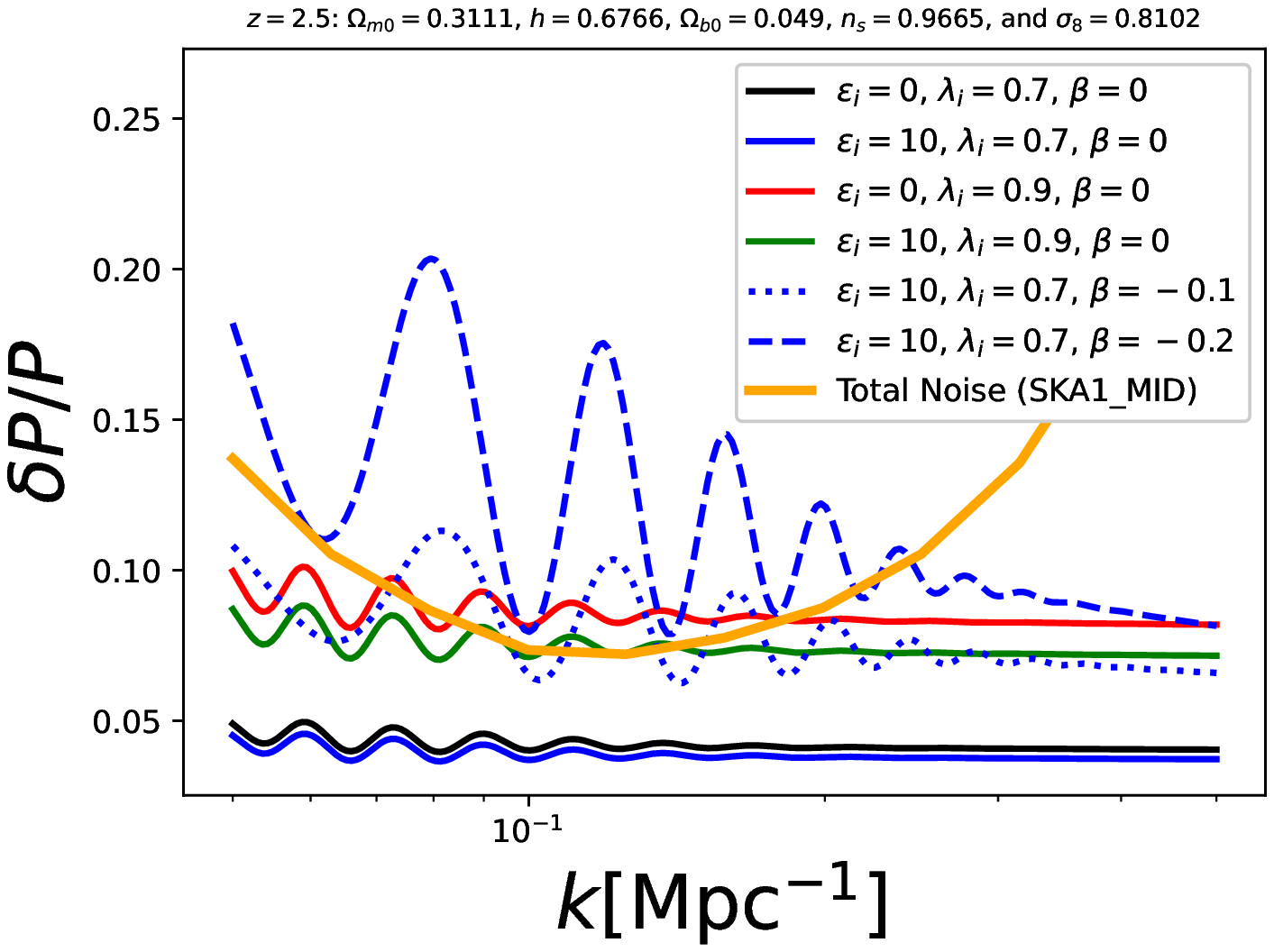}
\caption{
\label{fig:ps21D3LcdmSKA}
Detectability of interacting cubic Galileon model from the $\Lambda$CDM model. Here we have plotted $\delta P/P$, where $P=P_{\text{fiducial}}=P_{21}(\Lambda CDM)$ and $\delta P = | P_{21}^{3D}-P_{21}(\Lambda CDM) |$. Parameter values are same as in Figure~\ref{fig:ps21D3Lcdm}.
}
\end{figure}

\section{Detectibility with SKA1-mid telescope}
\label{sec-ska1mid}

Here we are considering the detectability of the cubic Galileon model from the $\Lambda$CDM model with the SKA1-mid telescope. To do so, we need to consider the errors that arise in the observed 21 cm power spectrum. We only consider two types of errors here. One is the system noise and another one is the sample variance. The system noise arises from the instrument. The cosmic variance arises from the finite sampling of modes. We ignore other errors like astrophysical residual foregrounds, effects from the ionosphere, etc. So, we consider an ideal instrumental detection where these errors are completely removed and the only errors present are the system noise and the sample variance. We closely follow the approach of \citep{noise_err_1} to estimate the noise.

We assume a circularly symmetric antenna distribution ($\rho_{ant}$) which is a function of $l$ only, where $l$ is the distance from the centre of the antennae distributions. For the SKA1-mid telescope antenna distributions, we follow the document \url{https://astronomers.skatelescope.org/wp-content/uploads/2016/09/SKA-TEL-INSA-0000537-SKA1_Mid_Physical_Configuration_Coordinates_Rev_2-signed.pdf}. The SKA1-mid telescope specifications have 133 SKA antennas with 64 MEERKAT antennas. The 2D baseline distribution is then given by \citep{noise_err_1}

\begin{equation}
\rho_{2D} (U,\nu_{21}) = B(\nu_{21}) \int_{0}^{\infty} 2 \pi l dl \hspace{0.1 cm} \rho_{ant} (l) \int_{0}^{2 \pi} d\phi \hspace{0.1 cm} \rho_{ant} (|\vec{l}-\lambda_{21} \vec{U}|),
\label{eq:rho2d}
\end{equation}

\noindent
where $\vec{U}$ is the baseline vector given by $\vec{U} = \frac{\vec{k}_{\perp} r}{2 \pi}$. $\vec{k}_{\perp}$ wavevector at the transeverse direction. $\nu_{21}$ and $\lambda_{21}$ are the observed frequency and the wavelength of the 21 cm signal respectively. $\phi$ is the angle between $\vec{l}$ and $\vec{U}$. $ B(\nu_{21})$ is computed by the normalization given by

\begin{equation}
\int_{0}^{\infty} U dU \int_{0}^{\pi} d\phi \hspace{0.1 cm} \rho_{2D} (U,\nu_{21}) = 1.
\label{eq:rho2dnorm}
\end{equation}

The 3D baseline distribution ($\rho_{3D}(k,\nu_{21})$) is defined as

\begin{equation}
\rho_{3D}(k,\nu_{21}) = \left[ \int_{0}^{1} d\mu \hspace{0.2 cm} \rho_{2D}^{2} \left( \frac{r k}{2 \pi} \sqrt{1-\mu^{2}},\nu_{21} \right) \right] ^{\frac{1}{2}},
\label{eq:rho3d}
\end{equation}

The system noise in the 21 cm power spectrum is given by \citep{noise_err_1,noise_err_2,noise_err_3,noise_err_4,Dinda:2018uwm,Hotinli:2021xln}

\begin{equation}
\delta P_{N} (k,\nu_{21}) = \frac{T_{sys}^{2}}{B t_{0}} \Big{(} \frac{\lambda_{21}^{2}}{A_{e}} \Big{)}^{2} \frac{2 r^{2} L}{N_{t} (N_{t}-1) \rho_{3D} (k,\nu_{21})} \frac{1}{\sqrt{N_{k}(k)}}.
\label{eq:pnfinal}
\end{equation}

\noindent
where $N_t$ is the total number of antennas (here 197). $B$ is the bandwidth corresponding to the observed signal and $L$ is the corresponding comoving length. $t_{0}$ is the observation time. $A_{e}$ is the effective collecting area of an individual antenna. This is related to the physical collecting area, $A$ of an antenna given by $A_{e}=e A $, where $e$ is the efficiency of an antenna. We consider $e=0.7$ and $A\approx 1256.6 m^2$ for an SKA1-mid antenna. $ T_{sys} $ is the system temperature. To compute system temperature, we closely follow ... $N_{k}$ is the total number of independent modes in the range between $k$ to $k+dk$ given by $N_{k}(k) = \frac{2 \pi k^{2} dk}{V_{1-mode}}$, where $V_{1-mode}$ is the volume occupied by a single independent mode given by $V_{1-mode} = \frac{(2 \pi)^{3} A}{r^{2} L \lambda_{21}^{2}}$.

The sample variance in 21 cm power spectrum is given by \cite{noise_err_1,noise_err_2,noise_err_3,noise_err_4,Dinda:2018uwm,Hotinli:2021xln}

\begin{equation}
\delta P_{SV} (k,\nu_{21}) = \Big{[} \sum_{\theta} \frac{N_{m}(k,\theta)}{P_{21}^{2}(k,\theta)} \Big{]}^{- \frac{1}{2}},
\label{eq:sv}
\end{equation}

\noindent
where $N_{m}(k,\theta) = \frac{2 \pi k^{2} dk \sin\theta d\theta}{V_{1-mode}}$. Thus, total noise is given by 

\begin{equation}
\delta P_{tot} (k,\nu_{21}) = \sqrt{ \delta P_{N}^2 (k,\nu_{21}) + \delta P_{SV}^2 (k,\nu_{21}) } .
\label{eq:totalN}
\end{equation}

In Figure~\ref{fig:ps21D3LcdmSKA}, we have shown that the cubic Galileon model can be detected from the $\Lambda$CDM model with the 21 cm power spectrum. The parameter values and the color codes are the same as in Figure~\ref{fig:ps21D3Lcdm}. The orange line is the total error in the 21 cm power spectrum with the SKA1-mid telescope specifications. The lines above this orange line are detectable by the SKA1-mid telescope.

\section{conclusion}
\label{sec-conclusion}

In this work, we consider the cubic Galileon model which is a special case of Horndeski gravity. Because of the Galileon shift symmetry, this model is ghost-free. We consider the thawing kind of behavior in this model using proper initial conditions. We consider both the interacting and the non-interacting cubic Galileon model. By interaction, we mean the interaction between the total matter and the Galileon field. The potential is linear in this model which naturally arises in this kind of model. We show the deviations of this model from the $\Lambda$CDM one both through the background and the perturbative quantities. Our main focus here is to detect this model through the 21 cm observations. For this purpose, we consider interferometric observations like SKA1-mid observations. We consider system noise and the sample variance according to this observation and by considering $\Lambda$CDM as the fiducial model. We ignore other errors in the 21 cm power spectrum by the assumption that we can completely remove them for an ideal observation. With this error, we show the possibility of the detection of the cubic Galileon model from the $\Lambda$CDM model. Our results show that with forthcoming SKA observations, we can put strong constraints on the modified gravity models like the interacting cubic Galileon model. Also with SKA observation at higher redshifts, there is a bright possibility to distinguish such kind of Galileon models from $\Lambda$CDM which is surely encouraging.

\section{Acknowledgements}
AAS acknowledges the funding from SERB, Govt of India under the research grant no: CRG/2020/004347.


\begin{thebibliography}{99}

\bibitem{Riess:1998cb} 
  A.~G.~Riess {\it et al.} [Supernova Search Team],
  Astron.\ J.\  {\bf 116}, 1009 (1998)
  [astro-ph/9805201].
  
\bibitem{Perlmutter:1998np} 
  S.~Perlmutter {\it et al.} [Supernova Cosmology Project Collaboration],
 Astrophys.\ J.\  {\bf 517}, 565 (1999)
  [astro-ph/9812133].
  
\bibitem{Ade:2015xua} 
  P.~A.~R.~Ade {\it et al.} [Planck Collaboration],
  Astron.\ Astrophys.\  {\bf 594}, A13 (2016)
  [arXiv:1502.01589 [astro-ph.CO]].



\bibitem{Copeland:2006wr} 
  E.~J.~Copeland, M.~Sami and S.~Tsujikawa,
  Int.\ J.\ Mod.\ Phys.\ D {\bf 15}, 1753 (2006)
  [hep-th/0603057].
  
\bibitem{Linder:2008pp} 
  E.~V.~Linder,
  Rept.\ Prog.\ Phys.\  {\bf 71}, 056901 (2008)
  [arXiv:0801.2968 [astro-ph]].
  
\bibitem{Silvestri:2009hh} 
  A.~Silvestri and M.~Trodden,
  Rept.\ Prog.\ Phys.\  {\bf 72}, 096901 (2009)
  [arXiv:0904.0024 [astro-ph.CO]].
  
\bibitem{Sahni:1999gb} 
  V.~Sahni and A.~A.~Starobinsky,
  Int.\ J.\ Mod.\ Phys.\ D {\bf 9}, 373 (2000)
  [astro-ph/9904398].
  

\bibitem{Martin:2012bt}
J.~Martin,
Comptes Rendus Physique \textbf{13}, 566-665 (2012)
doi:10.1016/j.crhy.2012.04.008
[arXiv:1205.3365 [astro-ph.CO]].


\bibitem{Zlatev:1998tr}
I.~Zlatev, L.~M.~Wang and P.~J.~Steinhardt,
Phys. Rev. Lett. \textbf{82}, 896-899 (1999)
doi:10.1103/PhysRevLett.82.896
[arXiv:astro-ph/9807002 [astro-ph]].

\bibitem{Vafa:2005ui}
C.~Vafa,
[arXiv:hep-th/0509212 [hep-th]].

\bibitem{Obied:2018sgi}
G.~Obied, H.~Ooguri, L.~Spodyneiko and C.~Vafa,
[arXiv:1806.08362 [hep-th]].

\bibitem{Andriot:2018wzk}
D.~Andriot,
Phys. Lett. B \textbf{785}, 570-573 (2018)
doi:10.1016/j.physletb.2018.09.022
[arXiv:1806.10999 [hep-th]].


\bibitem{Riess:2019cxk}
A.~G.~Riess, S.~Casertano, W.~Yuan, L.~M.~Macri and D.~Scolnic,
Astrophys. J. \textbf{876}, no.1, 85 (2019)
doi:10.3847/1538-4357/ab1422
[arXiv:1903.07603 [astro-ph.CO]].

\bibitem{Wong:2019kwg}
K.~C.~Wong, S.~H.~Suyu, G.~C.~F.~Chen, C.~E.~Rusu, M.~Millon, D.~Sluse, V.~Bonvin, C.~D.~Fassnacht, S.~Taubenberger and M.~W.~Auger, \textit{et al.}
Mon. Not. Roy. Astron. Soc. \textbf{498}, no.1, 1420-1439 (2020)
doi:10.1093/mnras/stz3094
[arXiv:1907.04869 [astro-ph.CO]].

\bibitem{Pesce:2020xfe}
D.~W.~Pesce, J.~A.~Braatz, M.~J.~Reid, A.~G.~Riess, D.~Scolnic, J.~J.~Condon, F.~Gao, C.~Henkel, C.~M.~V.~Impellizzeri and C.~Y.~Kuo, \textit{et al.}
Astrophys. J. Lett. \textbf{891}, no.1, L1 (2020)
doi:10.3847/2041-8213/ab75f0
[arXiv:2001.09213 [astro-ph.CO]].


\bibitem{Planck:2018vyg}
N.~Aghanim \textit{et al.} [Planck],
Astron. Astrophys. \textbf{641}, A6 (2020)
[erratum: Astron. Astrophys. \textbf{652}, C4 (2021)]
doi:10.1051/0004-6361/201833910
[arXiv:1807.06209 [astro-ph.CO]].


\bibitem{Copeland:2006wr}
E.~J.~Copeland, M.~Sami and S.~Tsujikawa,
Int. J. Mod. Phys. D \textbf{15}, 1753-1936 (2006)
doi:10.1142/S021827180600942X
[arXiv:hep-th/0603057 [hep-th]].


\bibitem{Clifton:2011jh}
T.~Clifton, P.~G.~Ferreira, A.~Padilla and C.~Skordis,
Phys. Rept. \textbf{513}, 1-189 (2012)
doi:10.1016/j.physrep.2012.01.001
[arXiv:1106.2476 [astro-ph.CO]].

\bibitem{deRham:2014zqa}
C.~de Rham,
Living Rev. Rel. \textbf{17}, 7 (2014)
doi:10.12942/lrr-2014-7
[arXiv:1401.4173 [hep-th]].

\bibitem{deRham:2012az}
C.~de Rham,
Comptes Rendus Physique \textbf{13}, 666-681 (2012)
doi:10.1016/j.crhy.2012.04.006
[arXiv:1204.5492 [astro-ph.CO]].

\bibitem{DeFelice:2010aj}
A.~De Felice and S.~Tsujikawa,
Living Rev. Rel. \textbf{13}, 3 (2010)
doi:10.12942/lrr-2010-3
[arXiv:1002.4928 [gr-qc]].

\bibitem{Dvali:2000hr}
G.~R.~Dvali, G.~Gabadadze and M.~Porrati,
Phys. Lett. B \textbf{485}, 208-214 (2000)
doi:10.1016/S0370-2693(00)00669-9
[arXiv:hep-th/0005016 [hep-th]].

\bibitem{Luty:2003vm}
M.~A.~Luty, M.~Porrati and R.~Rattazzi,
JHEP \textbf{09}, 029 (2003)
doi:10.1088/1126-6708/2003/09/029
[arXiv:hep-th/0303116 [hep-th]].

\bibitem{Nicolis:2004qq}
A.~Nicolis and R.~Rattazzi,
JHEP \textbf{06}, 059 (2004)
doi:10.1088/1126-6708/2004/06/059
[arXiv:hep-th/0404159 [hep-th]].

\bibitem{Nicolis:2008in}
A.~Nicolis, R.~Rattazzi and E.~Trincherini,
Phys. Rev. D \textbf{79}, 064036 (2009)
doi:10.1103/PhysRevD.79.064036
[arXiv:0811.2197 [hep-th]].

\bibitem{Deffayet:2009wt}
C.~Deffayet, G.~Esposito-Farese and A.~Vikman,
Phys. Rev. D \textbf{79}, 084003 (2009)
doi:10.1103/PhysRevD.79.084003
[arXiv:0901.1314 [hep-th]].

\bibitem{Vainshtein:1972sx}
A.~I.~Vainshtein,
Phys. Lett. B \textbf{39}, 393-394 (1972)
doi:10.1016/0370-2693(72)90147-5

\bibitem{Horndeski:1974wa}
G.~W.~Horndeski,
Int. J. Theor. Phys. \textbf{10}, 363-384 (1974)
doi:10.1007/BF01807638

\bibitem{Kobayashi:2011nu}
T.~Kobayashi, M.~Yamaguchi and J.~Yokoyama,
Prog. Theor. Phys. \textbf{126}, 511-529 (2011)
doi:10.1143/PTP.126.511
[arXiv:1105.5723 [hep-th]].

\bibitem{Chow:2009fm}
N.~Chow and J.~Khoury,
Phys. Rev. D \textbf{80}, 024037 (2009)
doi:10.1103/PhysRevD.80.024037
[arXiv:0905.1325 [hep-th]].

\bibitem{Silva:2009km}
F.~P.~Silva and K.~Koyama,
Phys. Rev. D \textbf{80}, 121301 (2009)
doi:10.1103/PhysRevD.80.121301
[arXiv:0909.4538 [astro-ph.CO]].

\bibitem{Kobayashi:2010wa}
T.~Kobayashi,
Phys. Rev. D \textbf{81}, 103533 (2010)
doi:10.1103/PhysRevD.81.103533
[arXiv:1003.3281 [astro-ph.CO]].

\bibitem{Kobayashi:2009wr}
T.~Kobayashi, H.~Tashiro and D.~Suzuki,
Phys. Rev. D \textbf{81}, 063513 (2010)
doi:10.1103/PhysRevD.81.063513
[arXiv:0912.4641 [astro-ph.CO]].

\bibitem{Gannouji:2010au}
R.~Gannouji and M.~Sami,
Phys. Rev. D \textbf{82}, 024011 (2010)
doi:10.1103/PhysRevD.82.024011
[arXiv:1004.2808 [gr-qc]].

\bibitem{DeFelice:2010gb}
A.~De Felice, S.~Mukohyama and S.~Tsujikawa,
Phys. Rev. D \textbf{82}, 023524 (2010)
doi:10.1103/PhysRevD.82.023524
[arXiv:1006.0281 [astro-ph.CO]].

\bibitem{DeFelice:2010pv}
A.~De Felice and S.~Tsujikawa,
Phys. Rev. Lett. \textbf{105}, 111301 (2010)
doi:10.1103/PhysRevLett.105.111301
[arXiv:1007.2700 [astro-ph.CO]].

\bibitem{Ali:2010gr}
A.~Ali, R.~Gannouji and M.~Sami,
Phys. Rev. D \textbf{82}, 103015 (2010)
doi:10.1103/PhysRevD.82.103015
[arXiv:1008.1588 [astro-ph.CO]].

\bibitem{Mota:2010bs}
D.~F.~Mota, M.~Sandstad and T.~Zlosnik,
JHEP \textbf{12}, 051 (2010)
doi:10.1007/JHEP12(2010)051
[arXiv:1009.6151 [astro-ph.CO]].

\bibitem{Deffayet:2010qz}
C.~Deffayet, O.~Pujolas, I.~Sawicki and A.~Vikman,
JCAP \textbf{10}, 026 (2010)
doi:10.1088/1475-7516/2010/10/026
[arXiv:1008.0048 [hep-th]].

\bibitem{deRham:2010tw}
C.~de Rham, G.~Gabadadze, L.~Heisenberg and D.~Pirtskhalava,
Phys. Rev. D \textbf{83}, 103516 (2011)
doi:10.1103/PhysRevD.83.103516
[arXiv:1010.1780 [hep-th]].

\bibitem{deRham:2011by}
C.~de Rham and L.~Heisenberg,
Phys. Rev. D \textbf{84}, 043503 (2011)
doi:10.1103/PhysRevD.84.043503
[arXiv:1106.3312 [hep-th]].

\bibitem{Hossain:2012qm}
M.~W.~Hossain and A.~A.~Sen,
Phys. Lett. B \textbf{713}, 140-144 (2012)
doi:10.1016/j.physletb.2012.06.016
[arXiv:1201.6192 [astro-ph.CO]].

\bibitem{Ali:2012cv}
A.~Ali, R.~Gannouji, M.~W.~Hossain and M.~Sami,
Phys. Lett. B \textbf{718}, 5-14 (2012)
doi:10.1016/j.physletb.2012.10.009
[arXiv:1207.3959 [gr-qc]].



\bibitem{LIGOScientific:2017vwq}
B.~P.~Abbott \textit{et al.} [LIGO Scientific and Virgo],
Phys. Rev. Lett. \textbf{119}, no.16, 161101 (2017)
doi:10.1103/PhysRevLett.119.161101
[arXiv:1710.05832 [gr-qc]].

\bibitem{LIGOScientific:2017zic}
B.~P.~Abbott \textit{et al.} [LIGO Scientific, Virgo, Fermi-GBM and INTEGRAL],
Astrophys. J. Lett. \textbf{848}, no.2, L13 (2017)
doi:10.3847/2041-8213/aa920c
[arXiv:1710.05834 [astro-ph.HE]].

\bibitem{LIGOScientific:2017ync}
B.~P.~Abbott \textit{et al.} [LIGO Scientific, Virgo, Fermi GBM, INTEGRAL, IceCube, AstroSat Cadmium Zinc Telluride Imager Team, IPN, Insight-Hxmt, ANTARES, Swift, AGILE Team, 1M2H Team, Dark Energy Camera GW-EM, DES, DLT40, GRAWITA, Fermi-LAT, ATCA, ASKAP, Las Cumbres Observatory Group, OzGrav, DWF (Deeper Wider Faster Program), AST3, CAASTRO, VINROUGE, MASTER, J-GEM, GROWTH, JAGWAR, CaltechNRAO, TTU-NRAO, NuSTAR, Pan-STARRS, MAXI Team, TZAC Consortium, KU, Nordic Optical Telescope, ePESSTO, GROND, Texas Tech University, SALT Group, TOROS, BOOTES, MWA, CALET, IKI-GW Follow-up, H.E.S.S., LOFAR, LWA, HAWC, Pierre Auger, ALMA, Euro VLBI Team, Pi of Sky, Chandra Team at McGill University, DFN, ATLAS Telescopes, High Time Resolution Universe Survey, RIMAS, RATIR and SKA South Africa/MeerKAT],
Astrophys. J. Lett. \textbf{848}, no.2, L12 (2017)
doi:10.3847/2041-8213/aa91c9
[arXiv:1710.05833 [astro-ph.HE]].

\bibitem{Ezquiaga:2017ekz}
J.~M.~Ezquiaga and M.~Zumalac\'arregui,
Phys. Rev. Lett. \textbf{119}, no.25, 251304 (2017)
doi:10.1103/PhysRevLett.119.251304
[arXiv:1710.05901 [astro-ph.CO]].

\bibitem{Zumalacarregui:2020cjh}
M.~Zumalacarregui,
Phys. Rev. D \textbf{102}, no.2, 023523 (2020)
doi:10.1103/PhysRevD.102.023523
[arXiv:2003.06396 [astro-ph.CO]].


\bibitem{Bartolo:2013ws}
N.~Bartolo, E.~Bellini, D.~Bertacca and S.~Matarrese,
JCAP \textbf{03}, 034 (2013)
doi:10.1088/1475-7516/2013/03/034
[arXiv:1301.4831 [astro-ph.CO]].

\bibitem{Bellini:2013hea}
E.~Bellini and R.~Jimenez,
Phys. Dark Univ. \textbf{2}, 179-183 (2013)
doi:10.1016/j.dark.2013.11.001
[arXiv:1306.1262 [astro-ph.CO]].

\bibitem{Barreira:2013eea}
A.~Barreira, B.~Li, W.~A.~Hellwing, C.~M.~Baugh and S.~Pascoli,
JCAP \textbf{10}, 027 (2013)
doi:10.1088/1475-7516/2013/10/027
[arXiv:1306.3219 [astro-ph.CO]].

\bibitem{Hossain:2017ica}
M.~W.~Hossain,
Phys. Rev. D \textbf{96}, no.2, 023506 (2017)
doi:10.1103/PhysRevD.96.023506
[arXiv:1704.07956 [gr-qc]].

\bibitem{Dinda:2017lpz}
B.~R.~Dinda, M.~Wali Hossain and A.~A.~Sen,
JCAP \textbf{01}, 045 (2018)
doi:10.1088/1475-7516/2018/01/045
[arXiv:1706.00567 [astro-ph.CO]].

\bibitem{Brahma:2019kch}
S.~Brahma and M.~W.~Hossain,
JHEP \textbf{06}, 070 (2019)
doi:10.1007/JHEP06(2019)070
[arXiv:1902.11014 [hep-th]].



\bibitem{Brahma:2020eqd}
S.~Brahma and M.~W.~Hossain,
Universe \textbf{7} (2021) no.6, 167
doi:10.3390/universe7060167
[arXiv:2007.06425 [astro-ph.CO]].


\bibitem{Renk:2017rzu}
J.~Renk, M.~Zumalac\'arregui, F.~Montanari and A.~Barreira,
JCAP \textbf{10} (2017), 020
doi:10.1088/1475-7516/2017/10/020
[arXiv:1707.02263 [astro-ph.CO]].


\bibitem{Wyithe:2007gz}
S.~Wyithe and A.~Loeb,
Mon. Not. Roy. Astron. Soc. \textbf{383} (2008), 606
doi:10.1111/j.1365-2966.2007.12568.x
[arXiv:0708.3392 [astro-ph]].

\bibitem{Loeb:2008hg}
A.~Loeb and S.~Wyithe,
Phys. Rev. Lett. \textbf{100} (2008), 161301
doi:10.1103/PhysRevLett.100.161301
[arXiv:0801.1677 [astro-ph]].


\bibitem{SKA:2018ckk}
D.~J.~Bacon \textit{et al.} [SKA],
Publ. Astron. Soc. Austral. \textbf{37} (2020), e007
doi:10.1017/pasa.2019.51
[arXiv:1811.02743 [astro-ph.CO]].


\bibitem[Braun et al.(2019)]{2019arXiv191212699B} Braun, R., Bonaldi, A., Bourke, T., et al.\ 2019, arXiv:1912.12699



\bibitem{riess}A.~G.~Riess et al., Astron.~J., {\bf 116}, 1009 (1998).

\bibitem{perlmutter}S.~Perlmutter et al., Astrophys.~J., {\bf 483}, 565 (1997).

\bibitem{spergel}D.~N.~Spergel et al., Astrophys.~J.~Suppl., {\bf 148}, 175 (2003).

\bibitem{hinshaw}G.~Hinshaw, Astrophys.~J.~Suppl., {\bf 148}, 135, (2003).

\bibitem{ade1}P.~A.~R.~Ade et al., Astron.~Astrophys., {\bf 594}, A13 (2016).

\bibitem{ade2}P.~A.~R.~Ade et al., Astron.~Astrophys., {\bf 594}, A14 (2016).

\bibitem{ade3}N.~Aghanim et al., Planck Collaboration (2018) [arXiv: astro-ph/1807.06209].

\bibitem{delubac}T.~Delubac et al., Astron.~Astrophys., {\bf 574}, A59 (2015).

\bibitem{ata}M.~Ata et al., MNRAS, {\bf 473}, 4773 (2018).

\bibitem{R16}A.~G.~Riess et al., Astrophys.~J., {\bf 826}, 56 (2016).

\bibitem{bonvin}V.~Bonvin et al., MNRAS, {\bf 465}, 4914 (2017).

\bibitem{valentino}E.~de,~Valentino, E.~Linder and A.~Melchiorri, Phys.~Rev.~D, {\bf 97}, 043528 (2018).

\bibitem{gongbo}Gong-bp~Zhao et al., Nat.Astron., {\bf 1}, 627 (2017).

\bibitem{sahni}V.~Sahni, A. Shafieloo and A.~A.~Starobinsky, Astrophys.~J., {\bf 793}, L40 (2014).



\bibitem{galBasic1} A. Nicolis, R. Rattazzi and E. Trincherini, The Galileon as a local modification of gravity, Phys. Rev. D 79 (2009) 064036 [arXiv:0811.2197] [INSPIRE].

\bibitem{galBasic2} N. Chow and J. Khoury, Phys. Rev. D 80, 024037 (2009) [arXiv:0905.1325 [hep-th]].

\bibitem{galBasic3} F. P. Silva and K. Koyama, Phys. Rev. D 80, 121301 (2009) [arXiv:0909.4538 [astro-ph.CO]].

\bibitem{galBasic4} T. Kobayashi, Phys. Rev. D 81, 103533 (2010) [arXiv:1003.3281 [astro-ph.CO]].

\bibitem{galBasic5} T. Kobayashi, H. Tashiro and D. Suzuki, Phys. Rev. D 81, 063513 (2010) [arXiv:0912.4641 [astro-ph.CO]].

\bibitem{galBasic6} A. De Felice, S. Mukohyama and S. Tsujikawa, Phys. Rev. D 82, 023524 (2010) [arXiv:1006.0281 [astro-ph.CO]].

\bibitem{galBasic7} C. Deffayet, O. Pujolas, I. Sawicki and A. Vikman, JCAP 1010, 026 (2010) [arXiv:1008.0048 [hep-th]].

\bibitem{galBasic8} C. de Rham, G. Gabadadze, L. Heisenberg and D. Pirtskhalava, Phys. Rev. D 83, 103516 (2011) [arXiv:1010.1780 [hep-th]].

\bibitem{galBasic9} C. de Rham and L. Heisenberg, Phys. Rev. D 84, 043503 (2011) [arXiv:1106.3312 [hep-th]].

\bibitem{galBasic10} L. Heisenberg, R. Kimura and K. Yamamoto, Phys. Rev. D 89, 103008 (2014) [arXiv:1403.2049 [hep-th]].

\bibitem{gal1} Nicola Bartolo, Emilio Bellini, Daniele Bertacca and Sabino Matarrese, Matter bispectrum in cubic Galileon cosmologies, JCAP03(2013)034.

\bibitem{gal2} Emilio Bellini and Raul Jimenez, The parameter space of cubic Galileon models for cosmic acceleration, Physics of the Dark Universe 2 (2013) 179$ - $183.

\bibitem{gal3} Barreira et. al., Nonlinear structure formation in the cubic Galileon gravity model, JCAP10(2013)027.

\bibitem{gal4} Bikash R. Dinda, Md. Wali Hossain and Anjan A Sen, Observed galaxy power spectrum in cubic Galileon model, [arxiv: astro-ph/1706.00567].

\bibitem{galfull8} M. A. Luty, M. Porrati, and R. Rattazzi, JHEP 09, 029 (2003), hep-th/0303116.

\bibitem{gal6} Md. Wali Hossain, First and second order cosmological perturbations in light mass Galileon models, Phys. Rev. D 96, 023506 (2017).

\bibitem{galfull1} Barreira et. al., The observational status of Galileon gravity after Planck, JCAP08(2014)059.

\bibitem{galfull2} C. Deffayet, G. Esposito-Farèse and A. Vikman, Covariant Galileon, Phys. Rev. D 79, 084003 (2009).

\bibitem{galfull3} R. Gannouji and M. Sami, Phys. Rev. D82, 024011 (2010), [arxiv:1004.2808].

\bibitem{galfull4} A. De Felice and S. Tsujikawa, Phys. Rev. Lett. 105, 111301 (2010), [arxiv:1007.2700].

\bibitem{galfull5} A. Ali, R. Gannouji, and M. Sami, Phys. Rev. D82, 103015 (2010), [arxiv:1008.1588].

\bibitem{galfull6} D. F. Mota, M. Sandstad, and T. Zlosnik, JHEP 12, 051 (2010), [arxiv:1009.6151].

\bibitem{galfull7} C. Deffayet, S. Deser, G. Esposito-Farese, Generalized Galileons: All scalar models whose curved background extensions maintain second-order field equations and stress tensors, Phys. Rev. D 80, 064015 (2009), [arxiv:0906.1967].



\bibitem{bharadwaj1}S.~Bharadwaj, B.~B.~Nath and S.~K.~Sethi, J.~Astrophys.~Astron., {\bf 22}, 21 (2001).

\bibitem{wyithe}J.~S.~B.~Wyithe and A.~Loeb, MNRAS, {\bf 397}, 1926 (2009).

\bibitem{bull} P.~Bull, P.~G.~Ferreira, P.~Patel and M.~G.~Santos, Astrophys.J., {\bf 803}, 21 (2015).

\bibitem{pade1}S. Capozziello, Ruchika and A. A. Sen, arXiv:1806.03943 [astro-ph.CO].

\bibitem{ska1}S.~Yahya, P.~Bull, M,~G.~Santos, M.~Silva, R.~Maartens, P.~Okouma and B.~Bassett, MNRAS, {\bf 450}, 2251 (2015).

\bibitem{ska2}M.~G.~Santos et al., Pos AASKA14 {\bf 019} (2015), arXiv:1501.04076 [astro-ph.CO].

\bibitem{ska3}R.~Maartens et al., Pos AASKA14 {\bf 016} (2015), arXiv:1501.03989 [astro-ph.CO].

\bibitem{quint}
C.~Wetterich, Nucl.\ Phys.\ B {\bf 302}, 668 (1988);
B.~Ratra and P.~J.~E.~Peebles, Phys.~Rev.~D, {\bf 37}, 3406 (1988);
P.~J.~E.~Peebles and B.~Ratra, Astrophys.~J, {\bf 325}, L17 (1988);
M. S. Turner and M. White, Phys. Rev. D {\bf{56}}, 4439 (1997); R. R. Caldwell,
R. Dave and P. J. Steinhardt, Phys. Rev. Lett. {\bf{80}}, 1582
(1998);  I.~Zlatev, L.~M.~Wang and P.~J.~Steinhardt, Phys.\ Rev.\
Lett.\ {\bf 82}, 896 (1999);

\bibitem{noncan}
  C. Armendariz-Picon, V. Mukhanov and P.J. Steinhardt, Phys. Rev. D, {\bf 63}, 103510 (2001);
  A.Y. Kamenschik, U. Moschella and V. Pasquier, Phys. Lett. B, {\bf 511}, 265 (2001);
  N. Billic, G.B. Tupper and R.D. Viollier, Phys. Lett. B, {\bf 535}, 17 (2002);
  M.C. Bento, O. Bertolami and A.A. Sen, Phys. Rev. D, {\bf 66}, 043507 (2002);
  T. Padmanabhan and T. Roy Choudhury, Phys. Rev. D, {66}, 081301 (2002);
  J. S. Bagla, H. K. Jassal and T. Padmanabhan, Phys. Rev. D {\bf 67}, 063504 (2003).
  A. Vikman, Phys. Rev. D, {\bf 71}, 023515 (2005);
  A. A. Sen, JCAP, {\bf 0603}, 10 (2006).
  
\bibitem{nonmin}
  J.P. Uzan, Phys. Rev. D, {\bf 59}, 123510 (1999);
  S. Sen and A. A. Sen, Phys. Rev. D, {\bf 63}, 124006 (2001);
  T.Chiba, Phys. Rev. D {\bf 60}, 083508) (1999);
  F. Perrotta, C. Baccigalupi, and S. Matarrese, Phys. Rev. D {\bf 61}, 023507 (2000);
  S. Nojiri and S. D. Odintsov and M. Sami, Phys. Rev. D {\bf 74}, 046004 (2006);
  V. Faraoni, Phys. Rev. D {\bf 68}, 063508 (2003).
  
\bibitem{phantom}
  R. R. Caldwell, Phys. Lett. B {\bf 545}, 23 (2002);
  E. Elizalde, S. Nojiri and S.D. Odintsov, Phys. Rev. D {\bf 70}, 043539 (2004);
  L. Revilolaropoulos, Phys. Rev. D {\bf 71}, 063503 (2005);
  T.Chiba, Phys. Rev. D {\bf 73}, 063501 (2006);
  J. Kujat, R.J. Scherrer and A. A. Sen, Phys. Rev. D {\bf 74}, 083501 (2006);
  R. J. Scherrer and A. A. Sen, Phys. Rev. D {\bf 78}, 067303 (2008).  
  
\bibitem{gal}A.~Nicolis, R.~Rattazzi and E.~Trincherini, Phys.~Rev.~D, {\bf 79}, 064036 (2009); 
  C.~Burrage, C.~de~Rham, L.~Heisenberg, JCAP, {\bf 1105}, 025 (2011);
  A.~Ali, R.~Gannouji and M.~Sami, Phys.~Rev.~D, {\bf 82}, 103015 (2010);
 R.~Gannouji and M.~Sami, Phys.~Rev.~D, {\bf 82}, 024011 (2010).
 
\bibitem{lindcald}R.~R.~Caldwell and E.~V.~Linder,
  Phys.\ Rev.\ Lett.\  {\bf 95}, 141301 (2005).
  
\bibitem{mog1} T. Clifton, P. G. Ferreira, A. Padilla and C. Skordis, Phys. Rept. 513, 1 (2012) [arXiv:1106.2476 [astro-ph.CO]].

\bibitem{mog2} K. Hinterbichler, Rev. Mod. Phys. 84, 671 (2012) [arXiv:1105.3735 [hep-th]].

\bibitem{mog3} C. de Rham, Comptes Rendus Physique 13, 666 (2012) [arXiv:1204.5492 [astro-ph.CO]].

\bibitem{mog4} C. de Rham, Living Rev. Rel. 17, 7 (2014) [arXiv:1401.4173 [hep-th]].

\bibitem{mog5} A. De Felice and S. Tsujikawa, Living Rev. Rel. 13, 3 (2010) [arXiv:1002.4928 [gr-qc]].
  
  
\bibitem{Saini:1999ba} 
  T.~D.~Saini, S.~Raychaudhury, V.~Sahni and A.~A.~Starobinsky,
  Phys.\ Rev.\ Lett.\  {\bf 85}, 1162 (2000)
  doi:10.1103/PhysRevLett.85.1162
  [astro-ph/9910231].
  

  
\bibitem{eisenhu}D.~J.~Eisenstein and W.~Hu, Astrophys.~J., {\bf 496}, 605 (1998).

\bibitem{PS_21_1} K.~ K.~Datta, T.~Roy~Choudhury and S.~Bharadwaj, MNRAS, {\bf 378}, 119 (2007).

\bibitem{PS_21_2} T.~Guha ~Sarkar, S.~Mitra, S.~Majumdar, T.~Roy~Choudhury, MNRAS, {\bf 421}, 3570 (2012).

\bibitem{PS_21_3} S.~Bharadwaj and S.~K.~Saiyad~Ali, MNRAS, {\bf 356}, 1519–1528 (2005).


\bibitem{PS_21_5} S.~R.~Furlanetto, M.~Zaldarriaga, and L.~Hernquist,  Astrophys.~ J, {\bf 613}, 16, (2004).

\bibitem{PS_21_6} A.~Hussain, S.~Thakur, T.~Guha Sarkar, and A.~A~Sen,  MNRAS {\bf 463}, 3493 (2016).

\bibitem{PS_21_7} A.~Liu, A.~R.~Parsons, MNRAS {\bf 457}, 1864 (2016).

\bibitem{PS_21_8} J.~Stuart,  B.~Wyithe and A.~Loeb,  MNRAS,  {\bf 397}, 1926 (2009).

\bibitem{PS_21_10} S.~K.~Saiyad Ali, S.~Bharadwaj and S.~K.~Pandey, MNRAS,  {\bf 366}, 213 (2006).

\bibitem{Bull} P.~Bull, P.~G.~Ferreira, P.~Patel, and M.~G.~Santos, Astrophys.~J., {\bf 803}, 21 (2015).

\bibitem{P21_3D_2} W.~E.~Ballinger, J.~A.~Peacock and A.~F. ~Heavens,  MNRAS, {\bf 282}, 877 (1996).

\bibitem{P21_3D_5} A.~Raccanelli et. al., arXiv:1501.03821 [astro-oh.CO].

\bibitem{noise_err_1} F.~Villaescusa-Navarro, M.~Viel, K.~K.~Datta and T.~Roy~Choudhury, JCAP, {\bf 09}, 050 (2014).

\bibitem{noise_err_2} T.~Guha~Sarkar and K.~K.~Datta,  JCAP,  {\bf 08} 001 (2015).

\bibitem{noise_err_3} P.~M.~Geil, B.~M.~Gaensler and J.~Stuart B.~Wyithe, MNRAS, {\bf 418}, 516 (2011).

\bibitem{noise_err_4} M.~McQuinn, O.~Zahn, M.~Zaldarriaga, L.~Hernquist, and S.~R.~Furlanetto, Astrophys.~J., {\bf 653}, 815 (2006).

\bibitem{Dinda:2018uwm}
B.~R.~Dinda, A.~A.~Sen and T.~R.~Choudhury,
[arXiv:1804.11137 [astro-ph.CO]].

\bibitem{Hotinli:2021xln}
S.~C.~Hotinli, T.~Binnie, J.~B.~Mu\~noz, B.~R.~Dinda and M.~Kamionkowski,
Phys. Rev. D \textbf{104} (2021) no.6, 063536
doi:10.1103/PhysRevD.104.063536
[arXiv:2106.11979 [astro-ph.CO]].

\end{thebibliography}
\end{document}